\documentclass[12pt,a4]{article}
\usepackage{amsmath}
\newcommand{\al}{\alpha}

\newcommand{\de}{\delta}

\newcommand{\eps}{\varepsilon}
\newcommand{\ga}{\gamma}
\newcommand{\Ga}{\Gamma}
\newcommand{\tGa}{\tilde{\Gamma}}
\newcommand{\ka}{\kappa}
\newcommand{\la}{\lambda}
\newcommand{\La}{\Lambda}
\newcommand{\Lao}{\Lambda_0}
\newcommand{\si}{\sigma}
\newcommand{\Si}{\Sigma}
\newcommand{\om}{\omega}

\newcommand{\vp}{\varphi}

\newcommand{\bc}{\bar{c}}
\newcommand{\uA}{\underline{A}}
\newcommand{\uB}{\underline{B}}
\newcommand{\uc}{\underline{c}}
\newcommand{\ubc}{\underline{\bar{c}}}
\newcommand{\uh}{\underline{h}}

\newcommand{\pa}{\partial}

\newcommand{\qed}{\hfill \rule {1ex}{1ex}\\ }
\newcommand{\eq}{\begin{equation}}
\newcommand{\eqe}{\end{equation}}
\newcounter{saveeqn}
\newcommand{\alpheqn}{\setcounter{saveeqn}{\value{equation}}
\setcounter{equation}{0}
\addtocounter{saveeqn}{1}
\renewcommand{\theequation}{\mbox{\arabic{saveeqn}\alph{equation}}}}
\newcommand{\reseteqn}{\setcounter{equation}{\value{saveeqn}}%
\renewcommand{\theequation}{\arabic{equation}}}

\newcommand{\Lf}{\mathcal{L}^{\Lambda ,\Lambda _0}}

\newcommand{\fit}{\varphi _{\tau }}
\newcommand{\fitx}{\varphi _{\tau }(x)}
\newcommand{\Jt}{J_{\tau }}
\newcommand{\Jtx}{J_{\tau }(x)}
\newcommand{\B}{B^{a}}

\newcommand{\jm}{j^{a}_{\mu}}

\newcommand{\jn}{j^{a}_{\nu}}
\newcommand{\ba}{b^{a}}

\newcommand{\ca}{c^{a}}
\newcommand{\cax}{c^{a}(x)}
\newcommand{\cb}{{\bar c}^{a}}
\newcommand{\cbx}{{\bar c}^{a}(x)}
\newcommand{\Am}{A^{a}_{\mu}}

\newcommand{\An}{A^{a}_{\nu}}
\newcommand{\et}{\eta ^{a}}
\newcommand{\etx}{\eta ^{a}(x)}
\newcommand{\etb}{{\bar \eta}^{a}}
\newcommand{\etbx}{{\bar \eta}^{a}(x)}
\newcommand{\g}{\gamma ^{a}}

\newcommand{\oma}{\omega ^{a}}

\newcommand{\LLz}{\Lambda, \Lambda_0}
\newcommand{\LLzz}{\Lambda_0, \Lambda_0}
\newcommand{\Lz}{\,0, \Lambda_0}
\newcommand{\sio}{\sigma _{0, \Lambda_0}}
\newcommand{\ufit}{\underline{\varphi }_{\tau}}
\newcommand{\uBa}{\underline B^{a}}
\newcommand{\uca}{\underline c^{a}}
\newcommand{\ucb}{\underline {\bar{c}}^{a}}
\newcommand{\uAn}{\underline A^{a}_{\nu}}
\newcommand{\uFi}{\underline \Phi }
\newcommand{\gt}{\gamma _{\tau}}

\newcommand{\Cll}{C^{\Lambda,\Lambda_0}}

\newcommand{\cLll}{{\cal L}^{\Lambda,\Lambda_0}}
\newcommand{\cLlln}{{\cal L}^{\Lambda,\Lambda_0}_{l,n}}

\begin{document}

\title{Renormalization of Spontaneaously Broken SU(2) Yang-Mills
Theory with Flow Equations}

\author{Christoph Kopper\footnote{\ kopper@cpht.polytechnique.fr} \\
Centre de Physique Th{\'e}orique, CNRS, UMR 7644\\
Ecole Polytechnique\\
F-91128 Palaiseau, France 
\and 
Volkhard F. M{\"u}ller\footnote{\ vfm@physik.uni-kl.de}\\
Fachbereich Physik, Technische Universit{\"a}t Kaiserslautern\\
D-67653 Kaiserslautern, Germany }

\date{ }

\maketitle

\begin{abstract}
Abstract: We present a renormalizability proof
for spontaneously broken $SU(2)$ gauge theory based on Flow
Equations. 
It is a conceptually and technically simplified version
of the earlier paper [KM] including some extensions.
The proof of  [KM] also was incomplete since an important
assumption made implicitly in the proof of Lemma 2 there 
is not verified. So the present paper is also a corrected
version of [KM].

\end{abstract}

\newpage

\section{Introduction }
The differential flow equations [WH] of the renormalization group 
 [W] offer a powerful tool for a unified approach to 
the analysis of systems with infinitely many degrees of freedom.
Although first conceived for an analysis of such systems beyond
perturbation theory, it was realized by  Polchinski [P]  
that these equations also paved the way for a new elegant approach
to  perturbative renormalization theory\footnote{Wilson himself remarked  
already in the late sixties 
that this should be possible,  as we learned from E. Br{\'e}zin.}.
Local gauge theories, however, present particular difficulties in
 this approach because the momentum space regulator violates gauge
invariance. Thus dimensional renormalization is in practice the most
popular scheme for renormalizing such theories in perturbation theory.
But at the same time this scheme is restricted to Feynman 
graphs. It not only defies to be given rigorous meaning in path integral
formulations, it does not  even directly apply in a mathematical sense 
to perturbative Green functions as a whole without splitting 
them into graphs. Thus, in some sense it is farthest away
from nonperturbative analysis, and it does not 
allow to address a number of interesting  conceptual, mathematical 
and quantitative questions. 
The authors analysed spontaneously broken
SU(2)-Yang-Mills theory with flow equations in [KM]. 
This analysis was simplified in [M].
 In an endeavour to further simplify and clarify the analysis, which was 
also caused by lecturing on the subject several times,
we came across an error in [KM], which reappeared in [M] by quotation.
In fact Lemma 2 in  [KM] cannot be proven without an assumption
made implicitly in its proof, 
which did not take into account the presence of irrelevant
boundary terms in the bare action. These terms have been ``forgotten''   
because  the context of the proof had changed in the progress of our
work, after the Lemma had been written. 
Since we have found quite a number of
further simplifcations in the mean time, since the subject is
important in physics, and since a correction of [KM] required quite a
lot of changes, even if the line of argument stays the same, we preferred
to write a self-contained modern and (hopefully~!) mathematically 
correct version of our previous paper.    

The strategy of proof remains that of [KM]. 
The (ultraviolet) power  counting part of the flow equation
renormalization proof is universal and simple for all 
renormalizable theories. For gauge theories
 we have to show that gauge
invariance can be restored when the cutoffs are taken away. On the level
of the Green functions  (which are not gauge invariant) 
this  means  that we have to verify the Slavnov-Taylor identities (STI)
 of the theory. They  then allow to argue that 
physical quantities such as the S-matrix are gauge-invariant [Z].
On analysing the flow equations (FE) for a gauge theory one realizes that 
the restoration of the STI depends on the choice of the
renormalization conditions chosen and cannot be true in general.
More precisely, since gauge invariance is violated in the regularized
theory, the renormalization group flow will generally produce
nonvanishing contributions to all those relevant parameters of the
theory, which are forbidden by gauge invariance, e.g. a noninvariant
gauge field self\-coupling of the form $(\vec{A}^2)^2$. 
The question is then: Can we use the freedom in adjusting the
renormalization conditions such that the STI are nevertheless 
restored in the end? To answer this question a first observation
is crucial: The violation of
the STI in the regularized theory can be expressed through Green functions
carrying an operator insertion, which depends on the regulators. FE theory
for such insertions tells us that these Green functions will vanish once the
cutoffs are removed, if we achieve renormalization conditions on
the noninserted Green functions 
such that the inserted ones, which are calculated from those, 
 have vanishing renormalization conditions
for all relevant terms, i.e. up to the dimension of the insertion 
(which is  5 in our case). Comparing the number of relevant terms for the
SU(2) theory - 37 (see App.A)-  and for the insertion 
- 53 (see App.C) -, we realize that it is not
possible to make vanish 53 terms on adjusting 37 free parameters,
unless there are linear interdependences. 
These interdependences are revealed in the analysis of the present
paper. As compared to [KM] we also include the proof of the validity
of the equation of the antighost in the renormalized theory
for suitable renormalization conditions.

This paper is organized as follows. In Section 2 we 
introduce the classical action of the model and
 the BRST-transformations, [BRS], [T].
In Section 3  we introduce the concepts from FE theory and recall the
statements on renormalizability we need. 
In particular we introduce the above mentioned operator
insertions.
When using FE it is natural to analyse the generating functional
of free propagator amputated Schwinger functions. 
The analysis of the STI is however technically simpler for 
one-particle irreducible vertex functions so that we
introduce the generating  functionals of both,
 together with the corresponding renormalizability statements.
In Section 4 we derive the violated Slavnov-Taylor identities (VSTI) 
for the regularized theory in various forms 
for the bare and the renormalized functionals. The Sections 1 to 4 
 follow closely the line of [KM]. 
In Section 5 we present the new tool required in view of the fact
that Lemma 2 of [KM] has become obsolete. Namely the generating
functional of the
vertex functions is not only expanded w.r.t. to fields and momenta,
but also w.r.t. the mass parameters, as far as their presence indicates 
improvement of UV power counting.  The corresponding redefinition
of relevant renormalization constants permits a {\it complete}   
analysis of the relevant part of the STI in terms of the renormalization
conditions. We do not need any more to jump from bare to renormalized
functionals and vice versa.
It is then possible to show that for suitable
renormalization conditions the inserted functional decribing the
violation of the STI has no relevant part. 
This result together with an obvious bound on its irrelevant part at the
regularization scale $\Lao$, following directly from the properties of
regulator, permits to prove that the violation disappears for   
$\Lao \to \infty$ so that the  STI hold in this limit.
This proof finally elucidates the fact the validity of the STI
can directly and fully be settled by analysing the (large) system of equations
describing its relevant part at the renormalization point.
This aim was not achieved in [KM]. 

We reproduce the appendices of [KM] with slight  notational changes.  
In Appendix A we list all 37 relevant terms allowed by the global
symmetries of $SU(2)$-Yang-Mills theory. In Appendix B the 7 relevant
 terms appearing in the inserted functionals describing the
 BRST-transformations are listed. In Appendix C we list 
the 53 equations corresponding to the relevant contributions
to the inserted functional describing the violation of the STI.
By analysis of this system of equations we show restoration of 
gauge symmetry in the (properly) renormalized theory.

A reader familiar with the
power counting results following from the flow equations 
can skip the major part of Section 3. He might  use it
for finding some notations also used in later Sections
and to get acquainted with the mass expansion of the Schwinger 
functions which is used  for the first time in this
paper. It is described in the last part of Section 3.1 
(from (\ref{scale}) onwards) and in the last page of Section 3.2
(from (\ref{nv1}) onwards).     
\section {The classical action}
Following closely the monograph of Faddeev and Slavnov [FS],
we collect some basic properties of the classical Euclidean
SU(2) Yang-Mills-Higgs model on four-dimensional Euclidean space-time.
The fields of the model are a triplet $\{A^a_{\mu}\}_{a =1,2,3}$ 
of real vector fields and the complex scalar doublet 
 $\{\phi_{\alpha}\}_{\alpha =1,2}$ .
The classical action has the form
\begin{equation}\label{y1}
S_{inv} = \int dx \left\{ \frac14 F^a_{\mu\nu} F^a_{\mu\nu} +
\frac12(\nabla_{\mu}\phi)^{\ast}
\nabla_{\mu}\phi + \lambda(\phi^{\ast}\phi-\rho^2)^2\right\},
\end{equation}
with the field strength tensor   
\begin{equation}\label{y2}
F^a_{\mu\nu}(x) = \partial_{\mu}A^a_{\nu}(x) -
\partial_{\nu}A^a_{\mu}(x) + g\epsilon^{abc} A^b_{\mu}(x) A^c_{\nu}(x)
\end{equation}
and the covariant derivative
\begin{equation}\label{y3}
\nabla_{\mu} = \partial_{\mu} + g \frac{1}{2i} \,\sigma^a A^a_{\mu}(x)
\end{equation}
acting on the SU(2)-spinor $\phi$. The parameters $g, \lambda, \rho$
are real positive, $\epsilon^{abc}$ is totally skew symmetric,
$\epsilon^{123} = +1$, and $\{\sigma^a\}_{a = 1,2,3}$ are the standard
Pauli matrices. The action (\ref{y1}) is invariant
under local gauge transformations of the fields  
\begin{equation}
\label{y4}
\begin{split}
&\frac{1}{2i}\, \sigma^a A^a_{\mu}(x) \longrightarrow u(x) \frac{1}{2i}\,
\sigma^a A^a_{\mu}(x) u^{\ast}(x) + g^{-1} u(x)
\partial_{\mu}u^{\ast}(x), \\ & \qquad \quad \phi(x) \longrightarrow
u(x) \phi(x), \quad\quad
\end{split}
\end{equation}
with $u: \mathbf{R}^4 \to $ SU(2), smooth.
The choice of a stable equilibrium point of
 the action (\ref{y1}) leads to spontaneous symmetry breaking, dealt
 with by reparametrizing the complex scalar doublet as
\begin{equation}
\phi(x) = \left( \begin{array}{c}
B^2(x) + i B^1(x) \\ \rho + h(x) - i B^3(x) \end{array} \right)\ ,
\label{y5}
\end{equation}      
  where $\{B^a(x)\}_{a=1,2,3}$ is a real triplet and $h(x)$ the real
Higgs field. Moreover, in place of the parameters $\rho, \lambda$ the
masses
\begin{equation}
m = \frac12 \,g \rho, \quad M = (8 \lambda \rho^2)^{\frac12}
\label{y6}
\end{equation}
are used. Aiming at a quantized theory, pure gauge degrees of
freedom have to be eliminated. We choose the  't Hooft gauge fixing,
with $\alpha \in \mathbf{R}_+$, 
\begin{equation}
S_{g.f.} = \frac{1}{2\alpha} \int dx (\partial_{\mu}A^a_{\mu} - \alpha
m B^a)^2\ .
\label{y7}
\end{equation}
  With regard to functional integration this condition is
implemented by introducing anticommuting Faddeev-Popov ghost and
antighost fields $\{c^a\}_{a=1,2,3}$ and $\{ \bar{c}^a\}_{a=1,2,3}\,$,
respectively, and forming with these six independent scalar fields the
additional term in the action 
\begin{equation}\label{y8}
 S_{gh} = - \int dx
    \bar{c}^a \big\{ (-\partial_{\mu}\partial_{\mu} + \alpha m^2)
      \delta^{ab} + \frac12 \,\alpha gmh \delta^{ab} 
      +\frac12\, \alpha gm \epsilon^{acb} B^c 
    - g \partial_{\mu} \epsilon^{acb} A^c_{\mu} \big\} c^b.
\end{equation} 
Hence, we have the total ``classical action" 
\begin{equation}
S_{\rm BRS} = S_{\rm inv} + S_{\rm g.f.} + S_{\rm gh} ,
\label{y9a}
\end{equation}
which is decomposed as
\begin{equation}
S_{\rm BRS} = \int dx \left\{ {\cal L}_{\rm quad} (x) + {\cal L}_{\rm
int}(x) \right\}
\label{y9b}
\end{equation}    
into its quadratic part, where $\Delta \equiv \partial_{\mu}
\partial_{\mu}\,$,
\begin{eqnarray}\label{y10}  
{\cal L}_{\rm quad}&=& \frac14 \,( \partial_{\mu}A^a_{\nu} -
\partial_{\nu}A^a_{\mu} )^2 + \frac{1}{2\alpha}\,
(\partial_{\mu}A^a_{\mu})^2 + \frac12\, m^2A^a_{\mu}A^a_{\mu} 
      + \frac12\, h (- \Delta + M^2) h   \nonumber \\
  &  & + \, \frac12 \,B^a (- \Delta + \alpha m^2) B^a 
        - \bar{c}^a (- \Delta + \alpha m^2) c^{a} \,,
\end{eqnarray}
and into its interaction part 
\begin{eqnarray}
{\cal L}_{\rm int} &=& g \epsilon^{abc}
(\partial_{\mu}A^a_{\nu})A^b_{\mu} A^c_{\nu} +
\frac14 \,g^2(\epsilon^{abc} A^b_{\mu}A^c_{\nu})^2 \nonumber \\
& & +\, \frac12 \,g \left\{ (\partial_{\mu}h)A^a_{\mu}B^a 
    - h A^a_{\mu}\partial_{\mu}B^a -
\epsilon^{abc}A^a_{\mu}(\partial_{\mu}B^b) B^c \right\} \nonumber \\
& & +\, \frac18\, g A^a_{\mu}A^a_{\mu} \left\{ 4 mh + g(h^2+B^aB^a)
               \right\} \nonumber \\
 & & +\, \frac14\, g \frac{M^2}{m} h (h^2 + B^aB^a) +
\frac{1}{32}\, g^2 \left( \frac Mm \right)^2 (h^2+B^aB^a)^2 \nonumber \\
  & &  - \, \frac12 \,\alpha gm \bar{c}^a \left\{ h \delta^{ab} +
\epsilon^{acb}B^c \right\} c^b
 - g \,\epsilon^{acb} (\partial_{\mu} \bar{c}^a) A^c_{\mu}c^b\ .
\label{y11}
\end{eqnarray}
Inspecting the quadratic part (\ref{y10}) we recognize two
 favourable consequences
 of the particular gauge fixing (\ref{y7}) : this part
 is diagonal in the fields 
 (no coupling $A^a_{\mu}\partial_{\mu} B^a $ appears) and 
all fields are massive. \\ 
As a prerequisite to state the symmetries of $S_{BRS}$
 (\ref{y9b}), composite
classical fields are introduced as follows:
\begin{equation} \label{y12}
\begin{split}
\psi^a_{\mu}(x)& = \left\{ \partial_{\mu} \,\delta^{ab} +
 g \epsilon^{arb} A^r_{\mu}(x)\right \}c^b(x), \\
\psi(x) & = - \frac12 \,g B^a(x) c^a(x),   \\
\psi^a(x) &= \big \{ (m + \frac12\, g\,h(x)) \delta^{ab} +
 \frac12 \,g \epsilon^{arb} B^r(x) \big \}c^b(x),  \\
\Omega^a(x) &= \frac12\, g \epsilon^{apq} c^p(x)c^q(x)\ . 
\end{split}
\end{equation}
We can then write (\ref{y8}) in the form
\begin{equation}
S_{gh} = - \int dx \, \bar{c}^a \{-\partial_{\mu}\psi^a_{\mu} +
          \alpha m \psi^a \}\ .
\label{y14}
\end{equation}
 The classical action $S_{BRS}$ (\ref{y9b}), shows the
following symmetries:
\begin{enumerate}
\item[i)] Euclidean invariance: $S_{BRS}$ is an O(4)-scalar.
\item[ii)] Rigid SO(3)-isosymmetry:
 The fields $\{A^a_{\mu}\}, \{B^a\}, \{c^a\}, \{ \bar{c}^a\}$
are isovectors and $h$ an isoscalar; $S_{BRS}$ is invariant under
spacetime independent SO(3)-transformations.
\item[iii)] BRS-invariance:\\
 The BRS-transformations of the basic fields [BRS] are defined as       
\begin{eqnarray} \label{y13}
A^a_{\mu}(x) &\longrightarrow &A^a_{\mu}(x) - \psi^a_{\mu}(x)\, \eps,
    \nonumber \\
h(x) &\longrightarrow &h(x) - \psi(x)\, \eps, \nonumber \\
B^a(x) &\longrightarrow& B^a(x) - \psi^a(x)\, \eps,   \\
c^a(x)& \longrightarrow& c^a(x) - \Omega^a(x)\,  \eps, \nonumber \\
\bar{c}^a(x)& \longrightarrow& \bar{c}^a(x) -
 \frac{1}{\alpha}\, \big ( \partial_{\nu}A^a_{\nu}(x)-
 \alpha m B^a(x) \big )\, \eps \nonumber
\end{eqnarray}
\end{enumerate}  
with the composite fields (\ref{y12}), and $\,\eps\,$ is a 
Grassmann element not depending on space-time, 
that commutes with the fields $\{A^a_{\mu}, h, B^a\}$
but anticommutes with the (anti-) ghosts $\{c^a, \bar{c}^a\}$.\\
To show the BRS-invariance of the total classical action
(\ref{y9a}) one first observes that the composite classical fields
(\ref{y12}) are themselves invariant under the BRS-transformations 
(\ref{y13}). Herewith, and using (\ref{y14}), it follows easily that 
the sum $ S_{\rm g.f.} + S_{\rm gh} $ is invariant under the 
transformation (\ref{y13}). Finally, on $ \,S_{\rm inv} $ act only 
the BRS-transformations of the fields $ A^a_{\mu} , B^a , h \,$,
 which amounts to local gauge transformations. \\
We observe that upon scaling the composite fields (\ref{y12}) 
entering the BRS-transformations as well as $S_{gh}$ (\ref{y14}),
 by a factor of $ \la $, the corresponding $ S_{\rm BRS} $ remains 
invariant under such BRS-transformations. 
\section{Renormalization without Slavnov-Taylor identities}
\subsection{ The Flow Equations for the Schwinger Functions}
 Quantization of the theory by means of functional integration 
  in the realm of (formal) power series 
 is based on a Gaussian measure
 related to the quadratic part (\ref{y10}) of $ S_{\rm BRS} $ (\ref{y9b}). 
 Denoting the differential operators appearing there by
 \begin{equation} \label{f1}
D_{\mu \nu} := (- \Delta +m^2 ) \,\delta_{\mu  \nu} -
 \frac{1-\alpha }{\alpha } \,\partial _{\mu}
  \partial _{\nu}\, ,\quad {\tilde D} := - \Delta + M^2 ,\quad
   D := -\Delta + \alpha  m^2 \,,
\end{equation} 
 we write
\begin{equation} \label{f2}
 \int dx\  {\mathcal L}_{\rm quad} (x) =
 \frac12 \,\langle A^{a}_\mu, D_{\mu  \nu }\, A^{a}_\nu \rangle 
 + \frac12 \,\langle h, {\tilde D} h \rangle
  +\frac12 \,\langle B^{a}, D B^{a} \rangle 
     - \langle {\bar c}^{\,a},  D c^{a} \rangle \,.
  \end{equation}
  To these differential operators (\ref{f1}) are
 associated the (free) propagators
  \eq \label{f3}
  C_{\mu \nu} (x, y) = \frac{1}{(2 \pi)^4} \, \int dk \
   e^{ik(x-y)}\  C_{\mu \nu} (k) \ ,
   \eqe
   and similarly in the other cases, with
    \begin{equation} \label{f4}
   C_{\mu  \nu}(k) = \frac{1}{k^2+m^2}
         \Big( \delta _{\mu \nu} - (1-\alpha )
      \frac{k_\mu k_\nu}{k^2+\alpha m^2} \Big) , \quad 
          C(k) = \frac{1}{k^2 +M^2} , \quad
    S(k) = \frac{1}{k^2 + \alpha m^2} \ .
  \end{equation}
A Gaussian product measure,
 the covariances of which are a regularized version
 of the propagators (\ref{f3}), (\ref{f4}), forms the point of departure.
 We choose the cutoff function, improving slightly the former one of [M],  
\begin{equation} \label{f5}
\sigma_{\Lambda}(k^2) \,=\,
 \exp \Big ( -  \frac{(k^2 + m^2)(k^2 + \alpha m^2)(k^2 + M^2) ( k^2)^2 }
{\Lambda^{10}} \,\Big )\ . 
\end{equation}
It is positive, invertible and analytic, and has the property 
 \begin{equation} \label{f6}
\frac{d}{d k^2} \,\sigma_{\Lambda}(k^2) |_{k^2 =0} \, = \, 0
 \end{equation} 
which will be helpful in the analysis of the relevant part of the STI
later on. 
Employing this cutoff function we define the regularized propagators,
 with  UV-cutoff $ \Lao < \infty \,$ and a flow parameter $\La$ 
satisfying
   $ 0 \leq \Lambda \leq \Lambda_0  $,
 \begin{equation} \label{f7}
 C^{\Lambda,\Lambda_0}_{\mu \nu}(k) \equiv  
   C_{\mu \nu}(k)\  \sigma_{\Lambda,\,\Lambda_0}(k^2)
   : = C_{\mu \nu}(k)\, \big( \,\sigma _{\Lambda_0}(k^2) -
  \sigma _{\Lambda}(k^2) \big) \,
 \end{equation}
 and similarly for  $ C(k), \, S(k)$. The
 particular choice (\ref{f5}) implies
 $$ \partial_{\La}\, C^{\Lambda,\Lambda_0}_{\mu \nu}(k) 
 = - \,\frac{10}{\Lambda^3}\cdot 
  \frac{ (k^2 + \alpha m^2)\delta_{\mu \nu} - (1-\al)k_{\mu}k_{\nu}}
    {\Lambda^2}\cdot \frac{(k^2+M^2) (k^2)^2}{\Lambda^6}
     \,\, \sigma_{\Lambda}(k^2) \,, $$
  and similarly in the other cases. Herefrom follow the bounds, 
   using $ C^{\Lambda,\Lambda_0}(k) $ as a
  collective symbol for the propagators considered, 
\begin{equation} \label{f8}
  {\big|} \partial^w
   \partial_{\Lambda}\, C^{\Lambda,\Lambda_0}(k) {\big |}
   \le \left\{ \begin{array}{ll} c_{|w|} \,
   \sigma_{2\Lambda}(k^2)
  & for \quad 0 \le \Lambda \le m\ , \\
 \Lambda ^{-3-|w|} P_{|w|}(\frac{|k|}{\Lambda}) \,\sigma _{\Lambda}(k^2) 
 & for \quad \Lambda > m\ . \end{array} \right\}
\end{equation}  
On the l.h.s. $ \pa^{\,w} $ denotes a $ |w|-$fold
 partial momentum derivative (see below (\ref{f23})).
 Moreover, the polynomials $ P_{|w|} $ have nonnegative coefficients, which,
 as well as the constants $ c_{|w|}$, depend on
$ \alpha, m, M, |w|$  only.
 Considering $ \si_{\La}(k^2) $, (\ref{f5}),
 as a function of $( \La , k^2 )$, it cannot be extended continuously 
to $ (0,0) $.
We set $ \si_{0}(0):= \lim_{k^2 \to 0} \si_{0}(k^2)= 0 $, and hence 
  $\si_{0,\Lao}(0)= \si_{\Lao}(0)= 1$.\\
 It is convenient to introduce   a short collective notation
for the various fields and their sources:
i) We denote the bosonic fields and the corresponding
 sources, respectively, by
\begin{equation} \label{f9}
\varphi _{\tau } = ( A^{a}_{\mu }\, , \, h,\, B^{a}) \
 ,\quad J_{\tau } = ( j^{a}_{\mu }\, , \, s, \,b^{a})\ ,
\end{equation}
ii) and all fields and their respective sources by
 \begin{equation} \label{f10}
\Phi  = ( \varphi _{\tau}\, , \, c^{a}, \, {\bar c}^{a})  \ , 
               \quad  K = ( J_{\tau}\, ,\, {\bar \eta}^{a},\, \eta ^{a})\ .
 \end{equation} 
 The sources $ \et\ $ and $ \etb $ are Grassmann elements and have ghost
 number $+1$ and $-1$, respectively. In the sequel, we exclusively use
  left derivatives with respect to these quantities.\\
  The characteristic functional of the Gaussian product measure
 with the covariances $\hbar C^{\,\Lambda, \Lambda_0} $
 from (\ref{f7}), (\ref{f4})
  is  then given by
 \begin{equation} \label{f11}
\int d\mu_{\LLz} (\Phi ) \, e^{\, \frac{1}{\hbar} \langle \Phi , K \rangle}
         \,=\  e^{\, \frac{1}{\hbar} P^{\Lambda, \Lambda_0}(K)}\ ,
 \end{equation}
 where
 \begin{equation} \label{f12}
\langle \Phi , K \rangle :\, = \int dx \Big( \sum_{\tau} \fitx \Jtx +
     \cbx \etx + \etbx \cax \Big)\ ,
 \end{equation}
 \eq \label{f13}
 P^{\La,\Lao}(K)  \,=\, 
    \frac12 \,\langle \jm, C^{\LLz}_{\mu \nu} \,\jn \rangle +
     \frac12 \,\langle s, C^{\LLz}\,  s \rangle 
     + \frac12\, \langle \ba , S^{\LLz} \,\ba \rangle 
           - \langle \etb, S^{\LLz}\,\et \rangle \ .
 \eqe 
 Aiming at a quantized descendant of the classical theory,
 we consider the
 generating functional $ L^{\LLz} (\Phi ) $ of the
 connected amputated Schwinger
 functions (CAS)
  \begin{eqnarray}
     \label{f14} 
     e^{-\frac{1}{\hbar}\left( L^{\Lambda ,\Lambda _0}(\Phi) 
               + I^{\Lambda ,\Lambda _0} \right)}
  & = & \int d\mu _{\Lambda ,\Lambda _0}(\Phi ' ) e^{-\frac{1}{\hbar}
                L^{\Lambda_0 ,\Lambda _0}( \Phi ' +\Phi ) }\ ,  \\ 
       \label{f15}
         L^{\Lambda , \Lambda _0}(0) & = & 0\ .
     \end{eqnarray}
The constant $ I^{\Lambda ,\Lambda _0} $ is the vacuum part
 of the theory which is proportional the volume
because of translation invariance. 
It therefore requires  to consider the theory at first in a finite 
   volume $ \Omega \subset \mathbf{R}^4 $. For details see [KMR].

 Since the regularization necessarily violates the local gauge symmetry,
 the bare functional
 \begin{equation} \label{f16}
   L^{\LLzz} (\Phi ) = \int dx \  \mathcal{L}_{\rm int} (x)\ +\
          L^{\LLzz}_{c.t.} (\Phi ) 
 \end{equation}
in a first stage has to be chosen sufficiently general in order to allow
for  the restoration of the Slavnov-Taylor
identities at the end. Therefore, we add to the interaction part (\ref{y11})
of classical origin  counter terms $ L^{\LLzz}_{c.t.} $, which a
priori include all local terms of mass dimension $\leq 4\, $
 permitted by the unbroken
global symmetries, i.e. Euclidean $O(4)$-invariance and
 $ SO(3) $-isosymmetry.  
     There are 37 such 
terms, by definition all at least 
of order $\mathcal{O}(\hbar) $. The general bare
 functional is presented in Appendix A. \\
 From (\ref{f14}) the corresponding flow equation follows
 upon differentiation with
 respect to the flow parameter $ \La\, $,\eq \label{f17}
  \pa_{\La} \,e^{-\frac{1}{\hbar}\left( L^{\Lambda ,\Lambda _0}(\Phi) 
               + I^{\Lambda ,\Lambda _0} \right)}
        = \hbar \,{\dot  \Delta _{\LLz}} \ 
          e^{-\frac{1}{\hbar}\left( L^{\Lambda ,\Lambda _0}(\Phi) 
               + I^{\Lambda ,\Lambda _0} \right)} \ ,
 \eqe  
where the r.h.s. is obtained on derivation of the  Gaussian measure
$ d{\mu}_{\LLz}(\Phi ')$ and observing that the integrand
is a function of $\,\Phi '+\Phi $.  
The ``dot'' appearing on the functional Laplace operator
  \eq \label{f18}
    \Delta _{\LLz} = \frac12 \,\big\langle
      \frac{\delta }{\delta \Am} , C^{\LLz}_{\mu \nu}
     \frac{\delta }{\delta \An}\big\rangle   
  + \frac12\,\big \langle \frac{\delta }{\delta h} , C^{\LLz}
       \frac{\delta }{\delta h} \big\rangle
  + \frac12\, \big \langle \frac{\delta }{\delta \B} ,
         S^{\LLz}\frac{\delta }{\delta \B} \big\rangle
    + \big \langle \frac{\delta }{\delta \ca} ,
         S^{\LLz}\frac{\delta }{\delta \cb} \big\rangle  
 \eqe
denotes differentiation with respect to $\La$.  Hence, we
 arrive at the flow equation
 \eq \label{flw}
 \begin{split}
 \pa_{\La} \left( L^{\Lambda ,\Lambda _0}(\Phi)
 + I^{\Lambda ,\Lambda _0} \right) =  \frac{\hbar}{2}\, \Big ( \sum_{\tau} 
        \Big \langle \frac{\delta }{\delta \fit} , 
  {\dot C}^{\LLz}_{\tau} \frac{\delta }{\delta \fit} \Big\rangle
 +2 \,\Big \langle \frac{\delta }{\delta \ca} ,
  {\dot S}^{\LLz}\frac{\delta }{\delta \cb} \Big\rangle  \Big )
     L^{\Lambda ,\Lambda _0}(\Phi) \\
 - \,\frac12\, \sum_{\tau} \Big \langle \frac{\delta L^{\LLz}}{\delta \fit} ,
  {\dot C}^{\LLz}_{\tau} \frac{\delta L^{\LLz}}{\delta \fit} \Big \rangle
     - \Big \langle \frac{\delta L^{\LLz}}{\delta \ca} ,
    {\dot S}^{\LLz}\frac{\delta L^{\LLz}}{\delta \cb} \Big \rangle \,. 
 \end{split}
\eqe 
 Since we restrict to perturbation theory, the generating functional
 will be considered within a formal loop expansion
  \begin{equation} \label{f19}
L^{\LLz}(\Phi ) =  \sum_{l=0}^{\infty} \hbar^l L^{\LLz}_{l}(\Phi)\ .
 \end{equation}  
Furthermore, decomposing into particular $n$-point Schwinger functions
we use a multiindex $n$, the components of which  denote the
 number of each source field species appearing:
 \begin{equation} \label{f20}
  n = ( n_A,\, n_h,\, n_B,\, n_{\bar c},\, n_c ) \ ,
    \quad |n| = n_A + n_h+ n_B + n_{\bar c} + n_c \ .
 \end{equation}  
Because of (\ref{y11}) there will not appear
 $1$- and $2$-point functions at the tree level ($l=0$).
 If we do not regard the vacuum part, we can study the
 flow of the $n$-point functions in the infinite volume limit
 $ \Omega \rightarrow \mathbf{R}^4 $.
 Due to translation invariance, it
 is convenient to consider also the
 Fourier transformed source field $ \hat \Phi $, the conventions used are
\begin{equation}\label{f21}
  \int_{p} := \int_{\mathbf{R}^4} \frac{d^{\,4} p}{(2\pi)^4}  \ ,\quad
  \Phi (x) = \int_{p} e^{ipx} \hat \Phi (p) \, \longrightarrow \,
  \delta _{\Phi (x)} := \frac{\delta }{\delta \Phi (x)}  
   = (2\pi )^4 \int_{p} e^{-ipx} \ \delta _{ \hat \Phi (p)} \,. 
\end{equation}
 Given these conventions, the momentum representation of the
 $n$-point function with multiindex $n$, (\ref{f20}),
 at loop order $l$ is obtained as an $|n|$-fold functional derivative
 \begin{equation} \label{f22}
  (2\pi )^{4(|n|-1)}\, \delta ^{\,n}_{\hat{\Phi }(p)} L^{\LLz}_l(\Phi) 
           |_{\Phi=0} \,
  = \, \delta (p_1 + \cdots + p_{|n|}) \,\Lf_{l,n} (p_1, \cdots , p_{|n|})\ .
 \end{equation}
 For the sake of a slim appearance, the notation does not reveal how
 the momenta are assigned to the multiindex $n$, 
 and in addition, the $ O(4) $- and $SO(3)$-tensor structure remains
 hidden. By definition the $n$-point function 
 is completely symmetric (antisymmetric) if the variables
 that belong to each of the bosonic (fermionic) species occurring are
  permuted. As momentum derivatives of  $ n $-point functions 
have to be considered, too, we also introduce the shorthand notation
\eq \label{f23}
w = (w_{1,1} , \cdots , w_{n-1,4} )\ , \quad w_{i,\mu } \in \mathbf{N}_0\  , 
  \quad \partial ^{\,w} : = \prod_{i=1}^{n-1} \prod_{\mu =1}^{4} 
  \Bigl(\frac{\partial }{\partial p_{i,\mu }} \Bigr)^{w_{i,\mu }} \, , 
 \quad |w| = \sum_{i,\mu } w_{i,\mu }\ .
 \eqe
The system of flow equations (FE) for the connected amputated Schwinger
functions (CAS) then follows 
from (\ref{flw}), using (\ref{f19}),(\ref{f22}), and finally 
performing the momentum derivatives (\ref{f23})    
\eq
\pa_{\La} \pa^w \,\cLlln (p_1, \cdots , p_{|n|}) \,=\,\!\!\!
\sum_{n',|n'|=|n|+2} c_{n-n'}   \int_k (\pa_{\La}\Cll(k))\,\pa^w 
\cLll_{l-1,\,n'}(k,-k,p_1, \cdots, p_{|n|})
\label{sflw}
\eqe
$$
-\!\!\!\!\!\!\!\!\!\!\!\sum_{l_1+l_2=l,\,w_1+w_2+w_3= 
 w\atop n_1,n_2,|n_1|+|n_2|=|n|+2} 
c_{\{w_i\}}\Biggl[ c_{n_1,n_2}\, \pa^{w_1} \cLll_{l_1,n_1}
 (p_1,\ldots,p_{|n_1|-1}, p')\qquad \qquad \qquad $$
$$\qquad \qquad  \qquad \qquad \qquad\cdot\, (\pa^{w_3}\pa_{\La}\Cll(p'))\,\,
 \, \pa^{w_2} \cLll_{l_2,n_2}(-p',\ldots,p_{|n|})\Biggr]_{s,a}\ .
$$
The field assignment of the propagators $\Cll$  on the r.h.s.
is not written, it is implicit in the multiindices
 $n',\ n_1, \ n_2$ related to $n$. In the
 linear term the integrated momentum $k$ refers to that
 of the fields from $n'-n$ and the factor $ c_{n-n'}$ has the value
 $ 1/2 $ and $1$ in the case of bosons and fermions, respectively.
 In the bilinear term we have $-p'= p_1+\ldots+p_{|n_1|-1}\,$.
 Furthermore the subscripts $s,a$ indicate full (anti)symmetrization
 according to the statistics of the various fields,
 requiring the combinatorial constants $ c_{n_1,n_2}$ 
 to rule out those permutations, which act solely within a 
 given CAS.\footnote{For details see [M], eq.(2.28).} 
The combinatoric coefficients $c_{\{w_i\}}\,$
stem form the Leibniz rule and have the values
$ c_{\{w_i\}}\,=\,\frac{w!}{w_1!w_2!w_3!}\,$, where
$w!\,=\,\prod_{i,\mu} w_{i,\mu}!\,\,$.

 To end up with Schwinger functions fulfilling the
 Slavnov-Taylor identities (STI), we have to consider 
Schwinger functions with a composite field inserted, too.
 Two kinds of such insertions have to be dealt with: local insertions
 implementing the BRS-variations, and a space-time integrated 
insertion representing the intermediate violation of the STI.\\
The classical composite BRS-fields (\ref{y12}) all have mass 
dimension $2$ and transform
as vector-isovector, scalar-isoscalar, scalar-isovector 
and scalar-isovector, respectively. Moreover, the
first three have ghost number $1$, whereas the last one 
has ghost number $2$. Hence, adding counterterms,
we introduce the bare composite fields 
\alpheqn
\begin{eqnarray}
\label{3.12a}
& & (\psi^a_{\mu})^{0,\Lao}(x) = R_1^{0}
\, \partial_{\mu} c^a(x) + R_2^{0}\, g\,
\epsilon^{arb}A^r_{\mu}(x)\,c^b(x)\ ,\\
\label{3.12b}
& & (\psi)^{0,\Lao}(x) = - R_3^{0} \,\frac12 g\, B^a(x) c^a(x)\ ,  
\\
\label{3.12c}
& & (\psi^a)^{0,\Lao}(x)  = R_4^{0} \, m \,c^a(x) 
+ R_5^{0} \,\frac12 g\,h(x)\,c^a(x)  + R_6^{0} \, \frac12 g\,
\epsilon^{arb} B^r(x) \,c^b(x)\ , 
\\
& &(\Omega^a)^{0,\Lao}(x)=  R_7^{0}
 \,\frac12 g\,\epsilon^{apq}c^p(x)c^q(x)\ ,
\label{3.12d}
\end{eqnarray}
\reseteqn
keeping the  notation from (\ref{y12}) but using
 it henceforth exclusively according to (\ref{3.12a})-(\ref{3.12d}). We set
 \begin{equation} \label{i2}
  R^0_{i} \, = \, 1+ \mathcal{O}(\hbar)\ , 
 \end{equation}
 thus viewing the counterterms again as formal power series in $\hbar $ ;
 the tree order ${\hbar}^0 $ provides the classical terms (\ref{y12}). 
 Observe that for $ l>0 $ the field products appearing in the
  classical composite fields $ \psi ^{a}_{\mu} $ and  $ \psi ^{a} $ of 
 (\ref{y12})  do require $ R^0_1 $ and $R^0_4$ , respectively,
 as counterterms.
 Moreover, it is important to note that the modified composite 
fields (\ref{3.12a})-(\ref{3.12d}) remain
 \emph{invariant} under the BRS-transformations (\ref{y13})
upon assuming the conditions
 \eq \label{i3}
 R ^0_6 = R^0_7 = R^0_2 \,, \qquad R^0_3\, R^0_5 = ( R^0_2 \,)^2
 \eqe
and employing the generalized composite fields
  (\ref{3.12a})-(\ref{3.12d})  in place of the   original ones, 
(\ref{y12}).\\
  To deal with Schwinger functions  showing \emph{one}
 insertion, the bare interaction (\ref{f16}) is modified adding 
 the composite fields (\ref{3.12a})-(\ref{3.12d}) coupled
 to corresponding sources
 \begin{equation} \label{i4}
\tilde {L}^{\LLzz}(\xi ; \Phi) : = L^{\LLzz}(\Phi) + L^{\LLzz} (\xi )\ ,
 \end{equation}
  \begin{equation} \label{i5}
 L^{\LLzz} (\xi ) = \int dx\ \{ \gamma^a_{\mu}(x) \psi^a_{\mu}(x) +
         \gamma(x) \psi(x)
 + \gamma^a (x)\psi^a(x)+ \omega^a(x) \Omega^a(x) \}\ . 
 \end{equation}
According to the properties of these composite fields, the
sources $\gamma^a_{\mu}, \gamma, \gamma^a$ are Grassmann elements,
they all have canonical dimension $2$ and ghost number $-1\,$, whereas
$\omega^a$ has canonical dimension $2\, $ and ghost number $-2$ .
For the insertions and their respective sources we also introduce a
short collective notation
\begin{equation} \label{i6}
   \psi _{\tau} = (\psi ^{a}_{\mu}\, , \, \psi , \, \psi ^{a} )
 \ , \quad  \gamma _{\tau} = ( \gamma ^{a}_{\mu}\, , \gamma ,
 \, \gamma ^{a}) \ ,
          \quad \xi = ( \gamma _{\tau}\, ,\, \omega ^{a})\ .
 \end{equation}
 Using now (\ref{i4}) in place of $ L^{\LLzz} $ as the bare action in 
 the representation (\ref{f14})  provides the functional 
 $ {\tilde L}^{\LLz}(\xi\,; \Phi)\, $, from which
 the generating functional of the regularized CAS
 with \emph{one}  insertion $\psi(x) $ follows as
 \begin{equation} \label{i7}
 L^{\LLz}_{\gamma }(x \, ; \Phi ) := \frac{\delta }{\delta \gamma (x)} 
          {\tilde L}^{\LLz}(\xi\,; \Phi)|_{\,\xi =0} \ , 
 \end{equation}
 and similarly for the other insertions from (\ref{i5}).
 In the infinite volume limit, and performing a Fourier
 transform of the insertion position we obtain
 \eq \label{i8}
 {\hat L}^{\LLz}_{\gamma }( q \, ; \Phi ) =
 \int dx \ e^{iqx}\ L^{\LLz}_{\gamma }(x \, ; \Phi ) \ .
 \eqe
 After loop expansion the $n$-point function
 with one insertion $\psi $ is obtained as
 \begin{equation} \label{i9}
    \delta ( q+p_1 + \cdots + p_{|n|})\ \Lf_{\gamma ; \,l, n}
             (q \, ;p_1, \cdots , p_{\,|n|}) :=
    (2\pi)^{4(|n|-1)} \delta ^{n}_{\hat{\Phi }(p)} 
{\hat L}^{\LLz}_{\gamma ; \, l}(q \, ;\Phi )|_{\Phi =0}  \ ,
 \end{equation}
 and similarly as regards the other insertions.\\
Starting from the analog of (\ref{flw}) for the modified
generating functional ${\tilde L}^{\LLz}(\xi ; \Phi)$, which
emerges from the bare action (\ref{i4}),
 and restricting to
one insertion by the operation (\ref{i7}), 
leads to a \emph{linear} flow equation
 for $ L^{\LLz}_{\gamma }(x \, ; \Phi )\,$. 
 Proceeding then as before in the derivation of (\ref{sflw}),
yields the system of differential FE for
the CAS  with\emph { one insertion $ \psi $}
\eq \label{sflwi}
\pa_{\La} \pa^w \,\Lf_{\ga;\, l, n}(q\,; p_1, \cdots , p_{|n|}) \,=\,\!\!\!
\sum_{n',|n'|=|n|+2}\!\! c_{n-n'}\int_k (\pa_{\La}\Cll(k))\,\pa ^w 
\cLll_{\ga;\,l-1,\,n'}(q;k,-k, p_1, \cdots , p_{|n|})
\eqe
\[
-
\!\!\!\!\!\!\!\!\!\sum_{l_1+l_2=l,\,w_1+w_2+w_3=w\atop
n_1,n_2,|n_1|+|n_2|=|n|+2} \!
c_{\{w_i\}}\,\Biggl[ c^{(1)}_{n_1,n_2}\, \pa^{w_1}
 \cLll_{\ga ;\,l_1,n_1}(q ; p_1,\cdots,p_{|n_1|-1}, p')
 \qquad \qquad \qquad \qquad
\]
\[
 \qquad \qquad \qquad \qquad \cdot \,(\pa^{w_3}\pa_{\La}\Cll(p'))\,
\pa^{w_2} \cLll_{l_2,n_2}(-p',\cdots,p_{|n|})\Biggr]_{s,a}\, .
\]
The notation is that of (\ref{sflw}), 
with $ - p' = q + p_1 + \cdots + p_{|n_1|-1} $, however.   
Since ghost and antighost in
(\ref{flw}) do not appear symmetrically, the 
$\bc$ ($c$)-derivative appears once
in $n_1$ ($n_2$) and once in $n_2$ ($n_1$). It is obvious that
each of the other insertions (\ref{i5})
leads to a similar system of flow equations.

 As will turn out in Section 4, the initial regularization,
 necessarily violating the STI, 
leads to a bare space-time integrated insertion of the form
\eq \label{i10}
L^{\LLzz}_1(\Phi )  =  \int dx\, N(x)\, ,  \qquad
 N(x)  =  Q(x) + Q '( x\, ; \Lambda_0^{-1} )\ . 
\eqe
The individual terms of $ N(x) $ involve at most five fields 
and have ghost number $1$. Furthermore, $Q(x)$ is a local 
polynomial in the fields and their derivatives,
 having  canonical mass dimension $ D=5$, whereas
 $ Q '( x \,; \Lambda_0^{-1}) $ is nonpolynomial in the 
field momenta but suppressed by powers of ${\Lambda_0}^{-1}\,$.
  To obtain the  generating functional $L^{\LLz}_1(\Phi )$ 
 with one (bare) insertion (\ref{i10}) we can resort to the local case,
 considering the bare local insertion
 \eq \label{i11}
  L^{\LLzz}(\varrho )  =   \int dx \, \varrho (x) N(x) 
 \eqe  
 and proceed as before. Observing (\ref{i7}), (\ref{i8}) we obtain
 \eq \label{i12}
   L^{\LLz}_{1} (\Phi )\,  =\, \int dx\ \frac{\delta }{\delta \varrho (x)} 
  {\tilde L}^{\LLz} (\varrho\,; \Phi )|_{\varrho =0}\, =\,
  \int dx\ L^{\LLz}_{\varrho} (x\,; \Phi )\, =\,
 {\hat L}^{\LLz}_{\varrho}(0; \Phi ) \ .
  \eqe
 Performing again a loop expansion, the CAS $n$-point function 
 with one insertion (\ref{i10}) is obtained as
 \begin{equation} \label{i13}
\delta ( p_1 + \cdots + p_{\,|n|})\
 {\mathcal L}^{\LLz}_{1; \, l, n}(p_1, \cdots , p_{|n|}) :=
(2 \pi)^{4(|n|-1)} \delta ^{\, n}_{{\hat \Phi }(p)}
 L^{\LLz}_{1; \, l}(\Phi ) |_{\, \Phi =0}\ . 
 \end{equation}
 For these CAS holds again a system of linear FE.  
 According to the preceding treatment of the integrated insertion 
 we only have to take (\ref{sflwi}) at the fixed momentum value
 $ q = 0 $ of the insertion,  
 and then replace each symbol
 $ \Lf_{\ga;\, l, n}(0\,; \cdots) $ by the new
 symbol $ {\mathcal L}^{\LLz}_{1; \, l, n}(\cdots )\,$.

Polchinski realized the flow equations (\ref{sflw}) to open the way for a
simple inductive proof of renormalizability. The mathematical proof
was carried through in [KKS] on simplifying still Polchinski's
argument. The FE for composite operators (\ref{sflwi})
 were introduced and analysed in [KK]. For a recent
presentation see [M]. 

The analysis of the STI, however, as will be shown in Section 4,
requires to trace in the perturbative expansion the
 effect of the super-renormalizable three-point couplings present
 in the interaction. To this end we scale in the tree-level
 part (\ref{y11}) of (\ref{f16}) the mass parameters appearing
 in the three-point couplings, as well as in the BRS-insertions
 the part proportional to $m\,$, see (\ref{3.12c}),
 by a common factor of $\la >0\,$:  
\eq
m \to \la m\ ,\quad
M \to \la M\  .
\label{scale}
\eqe
{\it Note however that we do not scale the mass parameters which are present
in the regularized propagators appearing in the flow equations.}
All CAS will  then depend smoothly 
on  $\,\la\,$, and we expand them as
\eq
{\mathcal L}^{\La,\Lao}_{l,n}(\la;\vec{p}) \,=\,
\sum_{\nu=0}^{\infty} (m\, \la)^{\nu}\
{\mathcal L}^{(\nu),\La,\Lao}_{l,n}(\vec{p}) \ ,\quad
 \vec{p}= (p_1,\cdots,p_{|n|})\, ,
\label{nl1}
\eqe
\eq
{\mathcal L}^{\La,\Lao}_{\ga ;\, l,n}(\la;q ;\vec{p}) \,=\,
\sum_{\nu=0}^{\infty} (m\,\la)^{\nu}\
{\mathcal L}^{(\nu),\La,\Lao}_{\ga;\,l,n}(q ;\vec{p}) \ ,
\label{nl2}
\eqe
where  for  suitable (physically natural~!) 
renormalization schemes the sum is finite,
its size depending on  $l\,$ and
$n\,$, as will be shown below.    
We adopt the following \\
{\it Renormalization scheme~:}
{\it Relevant terms} are those which satisfy\\
$|n|+|w|+\nu \le 4\,$ in case of the functional $\,L^{\LLz}\,$,
  $|n|+|w|+\nu \le 2\,$ in case of $\,L^{\LLz}_{\ga}\,$,\\
in agreement with the bounds to be derived below.\\ 
 At tree level we then have \footnote{Notice, that for
 $ l=0 $ there are no CAS with $ |n| \leq 2 $.} 
\eq
(\partial^w {\mathcal L}^{(\nu),\La,\Lao}_{0,n})(\vec{0})\,=\,0\,,
\quad \mbox{if }\quad |n|+|w|+\nu\, < \, 4\ .
\label{nl3}
\eqe
For $ l \geq 1$, we use renormalization
 conditions on the relevant terms as follows: we impose
\eq
(\partial^w {\mathcal L}^{(\nu),0,\Lao}_{l,n})(\vec 0) 
\stackrel{!}{=} \,0\,,
\quad \mbox{if }\quad |n|+|w|+\nu \, < \, 4  \ ,
\label{nl4}
\eqe
whereas if $\, |n|+|w|+\nu \, = \, 4 \,$, on the r.h.s. a free constant 
$ r_{(\nu),\, l,\, n}$  can be  chosen.\\
Correspondingly, in the case of an insertion, 
 we have at the tree level
\eq \label{nl5}
 (\partial^w {\cal L}^{(\nu), \La,\Lao}_{\ga ;\,0, n})(0; \vec{0}\,) = 0\, ,
\quad \mbox{if}\quad |n| + |w|+ \nu < 2 \,,
\eqe
and employ renormalization conditions
\eq \label{nl6}
 (\partial^w {\cal L}^{(\nu), 0 ,\Lao}_{\ga ;\,l,n})(0 ; \vec{0}\,)
   \stackrel{!}{=} 0 \, ,
 \quad \mbox{if}\quad  |n| + |w|+ \nu < 2 \,,
\eqe 
but if $\, |n|+|w|+\nu \, = \, 2 \,$, on the r.h.s. again
 a free constant can be  chosen.\\   
Because of the expansions (\ref{nl1}) and (\ref{nl2}) the FE
(\ref{sflw}) and (\ref{sflwi}) have to be adjusted attributing a
 superscript $ (\nu) $ to the CAS and to sum $ \nu_1 + \nu_2 = \nu $,
 in complete analogy to the loop index $l$. Using these extended
 FE the following bounds can be deduced,\\
\textbf{ Proposition 1}\\
\emph{Let $ l \in \mathbf{ N}_0 $ and $\, 0 \leq \La \leq \Lao $, then}
\eq
|\,\partial^w {\cal L}^{(\nu) , \La,\Lao}_{l,n}(\vec{p}\,)|
\,\leq\,
 (\La+m)^{4-|n|-|w|- \nu}\,{\cal P}_1(\log{\La+m \over m})\,
{\cal P}_2(\frac{|\vec{p}|}{\La+m})\ ,
\label{b1}
\eqe
\eq    
|\,\partial^w {\cal L}^{(\nu), \La,\Lao}_{\ga ;\,l,n}(q ; \vec{p}\,)| 
\,\leq\,
(\La+m)^{2-|n|-|w|- \nu}\,{\cal P}_1(\log{\La+m \over m})\,
{\cal P}_2(\frac{|q,\vec{p}|}{\La+m})\ .
\label{b2}
\eqe
\emph{ In these bounds ${\cal P}_i\,,\,i=1,2, $ denote (each time they
 appear possibly new) polynomials with nonnegative coefficients
 independent of $\La,\Lao,\, \vec p,\, q,\, m\,$. The
 coefficients may depend on $n,\,l,\, w,\,
$ and the other free parameters of the theory $\al,\,M/m\, ,\, g $.}\\
 These bounds are uniform in $\, \Lao $.
 The proof is solely based on power counting
for renormalizable theories, it does not involve
 the symmetry structure of the Yang-Mills theory.\\
\emph{Proof}: To prove (\ref{b1}) one  proceeds \emph{by induction}
  as follows: ascending in $ N:=2l+|n| $, for given $N$
  ascending in $ l $, for given $ N,l $ ascending in  $\nu $,
  and for given $ N, l, \nu $ descending in $|w| $.
  Given $\,n\,$, the irrelevant cases $ |n|+|w| + \nu > 4 $ 
  are treated first, integrating from
 the initial point $ \La = \Lao $ "downwards" with 
   initial conditions equal to zero. In contrast, the relevant ones,
 i.e.  $ |n|+|w| + \nu \leq 4 $, choosing the  particular 
  momentum value $ \vec p = 0 $, are integrated from the initial point
  $ \La = 0 $ "upwards" with initial conditions (\ref{nl4}) and the 
 remaining ones chosen freely, hereafter this result has to
 be extended to general $ \vec p \,$ via the Taylor formula
  $$ f(\vec p) = f(0) 
  + \vec p \cdot \int_{0}^{1} ({\vec \pa}\, f) (t \vec p \,)\, dt \, .$$
 Descending in $ |w| $, the integrand in the respective remainder
of the Taylor extension has already been bounded previously. 
 A derivative by induction  provides another factor of
$ \, (\La+m)^{-1}\, $,
 which can be combined with the momentum factor of the remainder
to increase the degree of the bounding  polynomial. A key to this
 induction is the property that in the tree order there
 are no CAS with $ | n| \leq 2 $. Bounding the linear term
 \footnote{This term generates a new loop.}  of the FE 
$$ \Big | \sum_{n',|n'|=|n|+2} c_{n-n'}   \int_k (\pa_{\La}\Cll(k))\,\pa^w 
 {\mathcal L}^{(\nu), \LLz}_{l-1,\,n'}(k,-k, {\vec p}\,)\Big | $$
$$ \leq 
\sum_{n',|n'|=|n|+2} \La\, \int_{k'} |\,{\La}^3\pa_{\La}\Cll(\La k')|
\,|\,\pa^w {\mathcal L}^{(\nu), \LLz} _{l-1,\,n'}(\La k',-\La k', {\vec p}\,)|
 $$
$$ \leq \La \sum_{n',|n'|=|n|+2} (\La+m)^{4-|n'|-|w|-\nu}
  \,{\cal P}_1(\log{\La+m \over m})\,
{\cal P}_2(\frac{|\vec{p}|}{\La+m}) $$
$$ \leq (\La+m)^{4-|n|-|w|-\nu - 1}
  \,{\cal P}_3(\log{\La+m \over m})\,
  {\cal P}_4(\frac{|\vec{p}|}{\La+m})\, , $$
 after a change of the integration variable $\, k=\La k' \,$
 one uses the bounds (\ref{f8}) and (\ref{b1}) and then 
 performs  the $ k'$-integration.\\
 The proof of (\ref{b2}) is analogous to the proof of (\ref{b1}):
 One has to observe the  inherent demarcation between relevant
 and irrelevant, and to employ the bound (\ref{b1}) required to treat
 the bilinear term on the r.h.s. of the FE (\ref{sflwi}) . 
\qed
Our renormalization scheme implies 
\eq
{\mathcal L}^{(\nu),\La,\Lao}_{l,n}(\vec{p})\equiv 0\ ,\
    \mbox{ if }\ \nu > 2l+|n|-2\ ,\quad
{\mathcal L}^{(\nu),\La,\Lao}_{\ga ;\,l,n}(q,\vec{p})\equiv 0\ ,\ 
 \mbox{ if }\ \nu > 2l+|n|-1\ .
\label{88}
\eqe
These statements follow inductively from the FE, once they hold for
the  terms fixed by the boundary conditions. 
Note that the first of these relations can be understood in
terms of Feynman graphs as following from the upper
bound on the number of trivalent vertices at a given loop-order.
The second one takes into account additionally  that the BRS-insertions
(\ref{3.12c}) also include one factor of $m\,$. 

 To also prove convergence
for $\Lao \to \infty\,$ (which a physicist would grant as a
consequence of uniformity) one has to analyse 
the FE, derived w.r.t. $\Lao\,$, 
using the same inductive technique. 
It is then possible to prove  [M]
that
\eq
|\partial_{\Lao}\partial^w {\cal L}^{(\nu),\,\La,\Lao}_{l,n}(\vec{p})|
\,\leq\,
\Lao ^{-2}\ (\La+m)^{5-|n|-|w|- \nu}\,{\cal P}_1(\log{\Lao \over m})\,
{\cal P}_2(\frac{|\vec{p}|}{\La+m})\ ,
\label{blao1}
\eqe
\eq
|\pa_{\Lao}\partial^w {\cal L}^{(\nu),\,\La,\Lao}_{\ga ;\,l,n}(q ; \vec{p})| 
\,\leq\,
\Lao ^{-2}\ (\La+m)^{3-|n|-|w|-\nu}\,{\cal P}_1(\log{\Lao \over m})\,
{\cal P}_2(\frac{|q,\vec{p}|}{\La+m})\ ,
\label{blao2}
\eqe
for $\Lao\,$ large  enough. Herefrom we can infer 
the existence of the limits 
$\Lao \to \infty\,$ at fixed value of $\La\,$.
%
%
%
\subsection{The Flow Equations for the Proper Vertex Functions}
Our analysis of the Slavnov-Taylor identities (STI) and the proof of 
their restoration will be based on a presentation in terms 
of proper vertex functions (1PI), since the extraction of relevant
parts from the STI is simpler and more transparent in terms of those
than in terms of the CAS. 
To present their relation with the CAS considered so far, we introduce
 the shorthand notation
 \eq \label{vs16}
  {\tilde L}(\xi; \Phi) := {\tilde L}^{\LLz}(\xi; \fit, c, \bar{c}) \ , 
\qquad C_{\tau} := C^{\LLz}_{\tau}\ ,
 \qquad S := S^{\LLz}\ , 
 \eqe  
 for the generating functional of the CAS with insertion (\ref{i5})
and for the  regularized propagators. 
From $\, {\tilde L}(\xi; \Phi)\, $ we define the
 "classical fields"  $\, \uFi \equiv (\ufit, \uc, \ubc) \, $ by
 \begin{equation} \label{vs17}
 \begin{split}
  \ufit (x) & =  \fit (x) - \int dy \,C_{\tau} (x-y) \,
                     \frac{\de {\tilde L}(\xi; \Phi)}{\de \fit(y)}\ , \\
  \uca (x) & =  \cax  +  \int dy \, S (x-y) \,
                      \frac{\de {\tilde L}(\xi; \Phi)}{\de \cb(y)} \ ,
 \quad \ucb (x) =  \cbx  - \int dy \, S (x-y) \,
                   \frac{\de {\tilde L}(\xi; \Phi)}{\de \ca(y)}\ .  
 \end{split} 
 \end{equation}
 The generating functional of the proper vertex functions
 $ \, {\tilde \Ga}( \xi ; \uFi) \equiv
 {\tilde \Ga}^{\LLz}(\xi; \ufit, \uc, \ubc) \,$ 
 is then given by the transform 
 \footnote{This transform corresponds to the familiar
  Legendre transform of the connected (non-amputated) Schwinger functions.}
 \eq \label{vs18}
  {\tilde \Ga}( \xi; \uFi)\, =\,{\tilde L}( \xi; \Phi) - \frac12 
   \sum_{\tau} \langle \fit, C^{ -1}_{\tau} \fit \rangle                 
               + \langle \bc, S^{-1} c \rangle 
         + \sum_{\tau} \langle \ufit C^{-1}_{\tau} \fit \rangle
    - \langle \bc, S^{-1} \uc \rangle - \langle \ubc, S^{-1} c \rangle \,,
  \eqe  
 with $ \Phi = \Phi (\uFi) $ on the r.h.s., according to (\ref{vs17}).
Since we are only interested in the kernels to be derived from
the generating functional $\Ga\,$ we may always assume the field
variables to be sufficiently regular so that the application of the
inverted regularized propagators makes sense.
By functional derivation we  deduce the relations
 \begin{equation} \label{vs19}
 \begin{split}
  \frac{\de {\tilde \Ga}( \xi; \uFi)}{\de \ufit(x)}  = &
                    \int dy\, C^{-1}_{\tau}(x-y)\, \fit(y) \ ,\\
   \frac{\de {\tilde \Ga}( \xi; \uFi)}{\de \uca(x)}  = &
                    \int dy\, S^{-1}(x-y)\, \cb(y) \ , \quad
   \frac{\de {\tilde \Ga}( \xi; \uFi)}{\de \ucb (x)}  = 
                           -  \int dy\, S^{-1}(x-y)\,\ca(y) \ ,
 \end{split}
 \end{equation}  
 forming the inverse of the relations (\ref{vs17}). Moreover, acting on
 the "classical fields" (\ref{vs17}) with the respective inverse
  propagators $\, C^{-1}_{\tau} \,$ and $ \, S^{-1} \,$,
 and then using (\ref{vs19}),
  provides the crucial relations between the generating
 functionals  $\, {\tilde L}(\xi; \Phi)\, $ and
 $ \, {\tilde \Ga}( \xi ; \uFi) \,$
  \begin{eqnarray} \label{vs20}
  (2\pi)^{-\,4}\, C^{-1}_{\tau}(p) \,\ufit(-p) & = &
   \frac{\de {\tilde \Ga}( \xi; \uFi)}{\de \ufit(p)} 
   \, - \, \frac{\de {\tilde L}( \xi; \Phi)}{\de \fit(p)}\,,\nonumber\\
    (2\pi)^{-\,4} S^{-1}(p) \,\uca(-p) & = &
   - \, \frac{\de {\tilde \Ga}( \xi; \uFi)}{\de \ucb(p)} 
   \, + \, \frac{\de {\tilde L}( \xi; \Phi)}{\de \cb(p)}\, , \\
    (2\pi)^{-\, 4} S^{-1}(p) \,\ucb(-p) & = &
    \frac{\de {\tilde \Ga}( \xi; \uFi)}{\de \uca(p)} 
   \, - \, \frac{\de {\tilde L}( \xi; \Phi)}{\de \ca(p)}\, , \nonumber
   \end{eqnarray}
 written in terms of Fourier transformed fields.
 Functional derivation of (\ref{vs18}) with respect to
 the source $ \ga(x) $ at fixed  $ \,\uFi \,$ leads to
  \eq \label{vs21}
    \frac{\de {\tilde \Ga}( \xi; \uFi)}{\de \ga(x)} {\Big|}_{\xi = 0}
   \, = \, \frac{\de {\tilde L}( \xi; \Phi)}{\de \ga(x)}
   {\Big |}_{\xi = 0} \ ,    
  \eqe  
  and to analogous equations as regards the other
  sources $ \ga^a_{\mu}, \,\ga^a, \, \om^a $.\\
 Restricting again to perturbation theory we
 consider the proper vertex functions which correspond to the various 
 types of CAS dealt with up to now. Hence, we define proper vertex
 functions without insertion, with one local insertion as in (\ref{i7}),
 (\ref{i8}), and with a global one as in (\ref{i12}), keeping the same
notations. Since by definition
 $ {\tilde \Ga}( \xi; \uFi) $ has no vacuum part, we
 can extend to infinite volume and use Fourier transformed "classical
 fields" (\ref{vs17}), with the conventions (\ref{f21})
 (but omitting the "hat" by abuse of notation). Hence, from the
 generating functionals $ \Ga^{\La,\Lao}_{l}\ , 
 \Ga^{\La,\Lao}_{\ga ;\, l} \ , \Ga^{\La,\Lao}_{1;\, l} $ we obtain the     
  corresponding $n$-point proper vertex functions of loop order $l$   
in analogy with  (\ref{f22}), (\ref{i9}), (\ref{i13}),  
\eq
(2 \pi)^{4(|n|-1)} \de^{n}_{\uFi(p)}\Ga^{\La,\Lao}_{l}(\uFi)
  |_{\uFi \equiv 0}\,=\,
\de(p_1+\dots+p_{|n|})\, \Ga^{\La,\Lao}_{l,n}(p_1,\cdots,p_{|n|})\ ,
\label{cag1}
\eqe
\eq
(2 \pi)^{4(|n|-1)} \de^{n}_{\uFi(p)}\Ga^{\La,\Lao}_{\ga\, ;\,l}(q;\,\uFi)
   |_{\uFi \equiv 0}\,=\,
\de(q+p_1+\dots+p_{|n|})\,\Ga^{\La,\Lao}_{\ga ;\,l,n}
         (q ; p_1,\cdots,p_{|n|})\ ,
\label{cag2}
\eqe
\eq
(2 \pi)^{4(|n|-1)} \de^{n}_{\uFi(p)}\Ga^{\La,\Lao}_{1;\,l}(\uFi)
   |_{\uFi\equiv 0}\,=\,
\de(p_1+\dots+p_{|n|})\, \Ga^{\La,\Lao}_{1;\,l,n}(p_1,\cdots,p_{|n|})\ .
\label{cag3}
\eqe

   The FE for the $ \tilde L $-functional implies a corresponding
  flow equation for the proper vertex functional $ \tGa $. Performing the
 $ \La $-derivative of the transform 
(\ref{vs18}) \footnote{again to be viewed on finite volume before
  passing to correlation functions} 
 and observing that the 
  classical fields $ \uFi\,$, (\ref{vs17}), themselves depend on $ \La\,$ 
  due to (\ref{vs19}),  eventually yields  
 \begin{eqnarray}\label{tfl}
( \pa_{\La} {\tilde \Ga}) ( \xi; \uFi)  = \,\pa_{\La} {\tilde L}( \xi; \Phi) 
 &-&\frac12  \sum_{\tau} \langle \fit,\pa_{\La} C^{ -1}_{\tau} \fit \rangle 
        + \langle \bc \,, \pa_{\La}S^{-1} c \rangle  \\
         &+& \sum_{\tau} \langle \ufit , \pa_{\La}C^{-1}_{\tau} \fit \rangle
    - \langle \bc\, , \pa_{\La} S^{-1} \uc \rangle 
               - \langle \ubc\, , \pa_{\La} S^{-1} c \rangle \ , \nonumber 
   \end{eqnarray}
where $\, (\pa_{\La} \tilde{\Ga})\,$ denotes the derivative
of the functional $\,\tilde{\Ga}\,$ itself. 
 Inserting now the flow equation for $ {\tilde L}( \xi; \Phi) $ which 
 has the same form as (\ref{flw}), and eliminating  in its bilinear
 terms the functionals  $ \frac{\de}{\de \Phi} \,{\tilde L} $ using
 the equations (\ref{vs17}),
  provides the flow equation of the vertex functional
   \eq \label{flwv}
 ( \pa_{\La} {\tilde \Ga}) ( \xi; \uFi)  + ( \pa_{\La} {\tilde I}) (\xi) 
 - \frac12 \sum_{\tau}\, \langle \ufit , \pa_{\La}C^{-1}_{\tau} \ufit \rangle
    + \langle \ubc \,, \pa_{\La} S^{-1} \uc \rangle 
        = \,\hbar \, {\dot \Delta}_{\LLz}\, {\tilde L}( \xi; \Phi) \ ,         
   \eqe
where one should remember  the dependence on  the
 parameters $ \LLz $  from (\ref{vs16}) and the definition (\ref{f18}).
At this stage the fields $ \uFi $ can be considered as autonomous
 (test) functions of the functional $ \tilde{\Ga}\,$, not depending on $\La$.
On the l.h.s. the second
 term is the vacuum part, since $ {\tilde \Ga} ( \xi; 0) = 0 $,
 and the subsequent terms  subtract  the (regularized) two-point tree
 order from  $ ( \pa_{\La} {\tilde \Ga}) ( \xi; \uFi)\, $. 
 The resulting functional still has  to be expressed in terms of
 proper vertex functions. Performing  a loop expansion and
functional derivatives w.r.t. the fields we obtain from  (\ref{flwv})
for $ \,|n| \geq 1 $
 \eq \label{flwv1}
  \de^{\, n}_{\uFi}|_{\uFi \equiv 0} : \, \,
  ( \pa_{\La} {\tilde \Ga}_{l})( \xi; \uFi)\,= \,
  {\dot \Delta}_{\LLz}\, {\tilde L}_{\, l-1}( \xi; \Phi)\, ,\quad l \geq 1\ . 
 \eqe
Since the vacuum part has disappeared we can now pass to  the infinite 
volume limit.
On the right hand side the functional
 $ {\tilde L}_{\, l-1}( \xi; \Phi)\,$ is first acted upon by two particular
  $ \Phi $-derivatives from the functional Laplace operator, then
   followed by an $n$-fold functional derivative with respect to (the 
   classical field) $ \uFi $. The resulting object
   has  to be expressed in terms of proper vertex functions. 
    There is no  closed formula for the r.h.s. in terms of proper
    vertex functions, and the presence of various types of fields 
    increases the combinatorial complexity. To indicate the
    procedure we employ a collective notation.
  We perform a $ \uFi$-derivative of the crucial relation  (\ref{vs20}),
    as regards $ {\tilde L}( \xi; \Phi)\,$ via the chain rule together
    with (\ref{vs19}), and hereafter consider the outcome within a
     loop expansion,
 \footnote{ Here $\Phi'''$ is determined by $\Phi''$, cf. (\ref{f2}).}
    \eq \label{clv}
    \frac {\de (p+q) \de_{l,\,0}}{ (2 \pi)^4 \, C_{\Phi, \Phi'}(p) } =
\frac{ \de^{\, 2} {\tilde \Ga}_{l}( \xi; \uFi)}{\de \uFi(p) \,\de \uFi '(q)}
      - (2\pi)^8 \sum_{\Phi '' \atop l_1 + l_2 = l} \int_{k}
 \frac{\de^{\,2} {\tilde L}_{l_1}( \xi; \Phi)}{\de \Phi(p) \,\de \Phi ''(k)}
   C_{\Phi '' \Phi'''}(k) \frac{ \de^{\, 2} {\tilde \Ga}_{l_2}( \xi; \uFi) }
         { \de \uFi'''(-k) \,\de \uFi '(q)}\,.
 \eqe
 This identity forms the point of departure to relate successively $n$-point
 functions of the $L$- and the $\Ga$- functional. We have to deal with it
  in the case without insertion, setting $\xi \equiv 0 $, as well as in the
 case of one local insertion. In the latter one, (\ref{clv}) has 
  to be derived with respect to the source at zero source,
 cf. (\ref{i7}),(\ref{i8}). By this operation, 
 both the $L$-functional with and without insertion appear,
 \begin{eqnarray} \label{clvi}
 \frac{ \de^{\, 2} {\Ga}^{\LLz}_{\ga ;\,l}( q ; \uFi) }
          { \de \uFi(p) \,\de \uFi '(p ')}
      &=& (2\pi)^8 \sum_{\Phi '' \atop l_1 + l_2 = l}\Big ( \,\int_{k}
 \frac{ \de^{\,2} {L}^{\LLz}_{\ga ;\,l_1}( q ; \Phi)}
        { \de \Phi(p) \, \de \Phi ''(k)}
     \, C^{\LLz}_{\Phi '' \Phi'''}(k) \frac{ \de^{\, 2} 
         {\Ga}^{\LLz}_{l_2}( \uFi) }
         { \de \uFi'''(-k) \,\de \uFi '(p ')} \nonumber \\
       & & \qquad \qquad + \int_{k}
 \frac{\de^{\,2} { L}^{\LLz}_{l_1}( \Phi)}{\de \Phi(p) \,\de \Phi ''(k)}
     \, C^{\LLz}_{\Phi '' \Phi'''}(k) \frac{ \de^{\, 2}
      {\Ga}^{\LLz}_{\ga ;\,l_2}( q ; \uFi) }
         { \de \uFi'''(-k) \,\de \uFi '(p ')} \,\Big ) \,.
 \end{eqnarray} 
 Taking (\ref{clvi}) at momentum $ q=0 $ and replacing the subscript $\ga$ 
 by the subscript $1$ provides the relation in the case of the
 integrated insertion.\\
 From (\ref{clv}) without insertion, considered at loop order
 $ l=0$ and at $ \Phi = \uFi \equiv 0 $, follows in the first step,
 because of the key property  
 $ {\mathcal L}^{\LLz}_{0 , n}(k, -k) \equiv 0 $, if $|n|=2 $,  
 \eq \label{clv1}
     1 =  C^{\LLz}_{\Phi, \Phi'}(p) \, \Ga^{\LLz}_{0,\, n}(p,-p) ,
         \quad n \,{\hat =} \,(\Phi, \Phi ') \ .
 \eqe  
 Before returning to the flow equation we note, that in order
 to obtain from (\ref{clv}) with $ \xi \equiv 0 $ or from (\ref{clvi})
 the relation between the various $n$-point functions of the $L$- 
 and the $\Ga$- functional, we have to act upon these equations
 repeatedly by $ \uFi $ - derivation, to be performed on the
  $L$-functional via the chain rule.
The chain rule derivatives $\de \Phi/\de {\uFi}\,$ can 
be read from (\ref{vs19}).   
In particular, on account of the propagators $C^{\LLz}(k)$
 vanishing at $ \La = \La_0 $ and observing (\ref{clv1}), one realizes,  
 ascending with $|n|$,
 \eq \label{clv2}
 \Ga^{\Lao,\Lao}_{l,n}(p_1,\cdots,p_{|n|})  =
  {\mathcal L}^{\Lao,\Lao}_{l,n}(p_1,\cdots,p_{|n|}) , 
           \quad (l, |n|) \,{\not =}\, (0, 2) \, ,     
 \eqe
 \eq \label{clv3}
   \Ga^{\Lao,\Lao}_{1;\,l,n}(p_1,\cdots,p_{|n|}) =
     {\mathcal L}^{\Lao,\Lao}_{1;\,l,n}(p_1,\cdots,p_{|n|}) \, ,
 \eqe 
 and similarly in the case of the local insertions.\\
 We now return to the FE (\ref{flwv1}) and first treat the case 
 without insertion, thus we set there $\xi \equiv 0$.
 Performing in addition the 
 momentum derivatives (\ref{f23}) we obtain the system, for
    $\, |n| \geq 1, \,(l, |n|) \,{\not =}\, (0,2) $,   
 \eq
 \pa_{\La} \pa^{\,w} \, \Ga^{\La,\Lao}_{l,n}(p_1, \cdots , p_{|n|}) \,=\,
 \frac12 \sum_{|n'|=|n|+2}' \int_k (\pa_{\La} C^{\La,\Lao}(k))\,\pa ^w 
 L^{\La,\Lao}_{l-1,\,n '}(k,-k; p_1, \cdots, p_{|n|})\, .
 \label{feg}
 \eqe
 The summation extends on the various propagators as stated in (\ref{clv}),
 not distinguished here notationally, the corresponding pair of 
fields together with $n$ determine $n'$. Moreover, the momentum 
 derivative $\pa^{\,w}$ concerns the momenta $ p_1, \cdots ,p_{|n|}$ of the 
 configuration $n$. To generate the functions on the 
 r.h.s. of (\ref{feg}) we have to act 
 on (\ref{flwv1}), after setting $\xi \equiv 0$, 
  with $\de^{\,n}_{\uFi}|_{\uFi \equiv 0}\,$, and these derivatives
are \emph{directly} applied on the $L$- functional.
  Hence the functions  $  L^{\LLz}_{l, n} $ 
 in (\ref{feg}), differing
  from  the CAS $\, {\mathcal L}^{\LLz}_{l, n}\, $. 
   The vanishing $2$-point CAS 
  in the tree order, together with its correspondence (\ref{clv1}) 
  then allow to express \emph{inductively} the  functions 
 $\, L^{\LLz}_{l, n}\, $ on the
  r.h.s. of (\ref{feg}) in terms of proper vertex functions,
  ascending in $l$, and for fixed $l$ ascending in $|n|\,$. 
The r.h.s of (\ref{feg}) then emerges in the form
 \eq \label{clv4}
  L^{\La,\Lao}_{l-1,\,n '}(k,-k; p_1, \cdots, p_{|n|})\, =
  {\Ga }^{\La,\Lao}_{l-1,\,n '}(k,-k, p_1, \cdots, p_{|n|})\,+ \cdots \, ,
  \eqe
 where the dots represent chains $\, \Ga \, C \, \Ga \,$ and 
higher iterations, formed of proper vertex
 functions $\, \Ga^{\LLz}_{l ',\, n ''}\,$ with
   $ (l ', n'') $ prior to $ (l-1, n ') $, joined via (free) propagators.\\
 In the case of one local insertion the equation (\ref{flwv1}) has
  to be derived with respect to the source at zero source,
 cf. (\ref{i7}),(\ref{i8}). Performing again the momentum derivation
 leads to the  the system of flow equations
  for proper vertex functions with one local insertion,
  $\, |n| \geq 1$, 
\eq
\pa_{\La} \pa^w \, \Ga^{\La,\Lao}_{\ga;\,l,n}(q ; p_1, \cdots, p_{|n|}) \,=\,
 \frac12\, \sum_{|n'|=|n|+2}'
 \int_k (\pa_{\La} C^{\La,\Lao}(k))\,\pa ^w 
 L^{\La,\Lao}_{\ga;\,l-1,n '}(q ; k,-k; p_1, \cdots, p_{|n|})\ ,
\label{fegi}
\eqe 
The r.h.s. of (\ref{fegi}) is now obtained in complete analogy
 to the case without
 insertion, the r.h.s. is now extracted inductively from
 (\ref{clvi}) in place of (\ref{clv}).
By this operation, both the $L$-functions with and without
  insertion appear. Proceeding inductively as before,
 and using the already determined $L$-functions without insertion,
 provides the function on the r.h.s. of the system (\ref{fegi}), as 
 \eq  \label{clv5}
  L^{\La,\Lao}_{\ga ;\, l-1,\,n '}(q;k,-k; p_1, \cdots, p_{|n|})\, =
  {\Ga }^{\La,\Lao}_{\ga;\, l-1,\,n '}(q;k,-k, p_1, \cdots, p_{|n|})\,
   + \cdots \, , 
\eqe
where the dots again represent a sum of chains,
 each of which contains exactly
 \emph{one} inserted factor $ \Ga^{\LLz}_{\ga ;\,l'', n''}$,
which has already been determined 
previously in the inductive procedure. \\
Finally, in the case of an integrated insertion, we obtain the
 system (\ref{fegi}) at the particular momentum value $ q \equiv 0 $. 

Once (\ref{clv4}) and (\ref{clv5}) have been inductively fixed, we can
again perform the mass scaling (\ref{scale}) in the tree-level 
interaction and insertions. It then leads to 
expansions corresponding  to  (\ref{nl1}), (\ref{nl2}) for the vertex
functions
\eq
{\Ga}^{\La,\Lao}_{l,n}(\la;\vec{p}\,) \,=\,
\sum_{\nu=0}^{\infty}(m\la) ^{\nu}\ 
{\Ga}^{(\nu),\La,\Lao}_{l,n}(\vec{p}\,) \ ,\quad
 \vec{p}= (p_1,\cdots,p_{|n|})\, ,
\label{nv1}
\eqe
\eq
{\Ga}^{\La,\Lao}_{\ga ;\, l,n}(\la;q ; \vec{p}\,) \,=\,
\sum_{\nu=0}^{\infty}(m\la) ^{\nu}\
{\Ga}^{(\nu),\La,\Lao}_{\ga;\,l,n}(q ; \vec{p}\,) \ .
\label{nv2}
\eqe
We first consider the tree level $ l=0$. In the case of
(\ref{nv1}) the scaling (\ref{scale}) of the interaction
 results in
\eq \label{nv3}
(\pa^{w}\,{\Ga}^{(\nu),0,\Lao}_{0, n})(\vec{0}\,) = 0 \,,
\quad |n| = 3,\quad |w| + \nu\, {\not =}\, 1\ .     
\eqe
Whereas there is no $ |n| = 1 $ content, the $2$-point 
functions are fixed by the regularized propagators (\ref{clv1})
(the masses of which are not scaled). The vertex functions
with insertion (\ref{nv2}) satisfy   
\eq \label{nv4}
(\pa^{ w}\,{\Ga}^{(\nu),0,\Lao}_{\ga ;\,0, n})(0 ; \vec{0}\,) = 0 \,,
\quad |n|+ |w| + \nu <  2\,.     
\eqe
Owing to the expansions (\ref{nv1}) and (\ref{nv2}), in both
 FE (\ref{feg}) and (\ref{fegi}) a superscript $ (\nu) $ has
to be attached to the respective $n$-point function on the
l.h.s. and on the $n'$-point functions present on the r.h.s.
We then use the same inductive scheme which leads to the bounds
(\ref{b1}),(\ref{b2}) on the CAS and may deduce
renormalizability of the proper vertex functions.
For the relevant terms the choice of the renormalization
conditions is as follows , $ l\geq 1\,$,
\eq \label{nv5}
(\pa^{\, w}\,{\Ga}^{(\nu),0,\Lao}_{l ,\, n})( \vec{0}\,)
 \stackrel{!}{=} 0 \,,
 \quad \mbox {if} \quad |n|+ |w| + \nu < 4\,,     
\eqe
but if $ |n|+ |w| + \nu = 4\,$, a nonvanishing 
 constant can be chosen on the r.h.s.,\\
 whereas in the case of an insertion
 \eq \label{nv6}
(\pa^{\, w}\,{\Ga}^{(\nu),0,\Lao}_{\ga ;\, l ,\, n})(0 ;\, \vec{0}\,)
 \stackrel{!}{=} 0 \,,
 \quad \mbox {if} \quad |n|+ |w| + \nu < 2\,,     
\eqe
but if $ |n|+ |w| + \nu = 2\,$, again a nonvanishing 
 constant on the r.h.s. may be imposed. \\
Proceeding inductively as indicated we obtain the bounds:\\
\textbf{ Proposition 2}\\
\eq
|\,\partial^w\, \Ga^{(\nu) , \La,\Lao}_{l,n}(\vec{p})|
\,\leq\,
 (\La+m)^{4-|n|-|w|-\nu }\,{\cal P}_1(\log{\La+m \over m})\,
{\cal P}_2(\frac{|\vec{p}|}{\La+m})\ ,
\quad (l, |n|)\, {\not =}\, (0, 2) \, ,
\label{bv1}
\eqe
\eq
|\,\partial^w \Ga^{(\nu), \La,\Lao}_{\ga ;\,l,n}(q ; \vec{p})| 
\,\leq\,
(\La+m)^{2-|n|-|w|- \nu}\,{\cal P}_1(\log{\La+m \over m})\,
{\cal P}_2(\frac{|q,\vec{p}|}{\La+m})\ ,
\label{bv2}
\eqe
\emph{The notations are those from (\ref{b1}),(\ref{b2}).}\\
Moreover, we can also obtain the bounds  (\ref{blao1}) - (\ref{blao2})
in the case of proper vertex functions derived w.r.t. $\Lao\,$.

\section{Violated Slavnov-Taylor identities}
 To examine the violation of the STI
produced by the UV cutoff $\Lambda_0$ we depart from
 the generating functional of the regularized Schwinger functions
 at the physical value $\Lambda = 0$ of the flow
 parameter, \footnote {Again one should stay in finite volume
  as long as the vacuum part is involved.} 
\eq   \label{vs1} 
   Z^{\Lz}(K) \, = \, \int d\mu_{\Lz}(\Phi)
   \,e^{-\frac{1}{\hbar} L^{\LLzz}(\Phi) 
     +\frac{1}{\hbar} \langle \Phi , K \rangle }\  .
\eqe 
 The Gaussian measure $ d\mu _{\Lz}(\Phi ) $ corresponds
to the quadratic form $ \frac{1}{\hbar}\, Q^{\Lz}(\Phi ) $,
cf. (\ref{f11}), 
\begin{equation} \label{vs2}
  Q^{\Lz}(\Phi )  = 
   \frac12 \langle A^{a}_\mu, \big( C^{\Lz}\big)^{-1}_{\mu  \nu }
        A^{a}_\nu \rangle 
  + \frac12 \langle h, (C^{\Lz})^{-1} h \rangle 
   +\frac12 \langle B^{a}, (S^{\Lz})^{-1} B^{a} \rangle -
  \langle {\bar c}^{a}, (S^{\Lz})^{-1} c^{a} \rangle .
\end{equation}  
 We define \emph{regularized} BRS-variations
 (\ref{y13}),(\ref{3.12a})-(\ref{3.12d})
 of the fields by
\begin{eqnarray} \label{vs3}
\delta _{BRS} \,\fitx & = & - \,
 ( \sio \,\psi _{\tau})(x) \,\eps ,\nonumber \\
\delta _{BRS} \,\ca (x) & = & - \, (\sio \,\Omega ^{a}) (x)\, \eps ,  \\
\delta _{BRS} \,\cb (x) & = & - \, \big (\sio \,(\, \frac{1}{\alpha }
 \,\partial _{\nu } \An -m \B)\big ) (x) \,\eps \ .\nonumber
 \end{eqnarray}  
  The BRS-variation of the Gaussian measure has the form 
\begin{equation} \label{vs4}
 d\mu _{\Lz}(\Phi ) \mapsto  d\mu _{\Lz}(\Phi ) \Big( 1- \frac{1}{\hbar}\,
 \delta _{BRS}\, Q^{\Lz}(\Phi ) \Big)  \,,
\end{equation}
 and inspecting (\ref{vs2}) we observe that the factor $\sio $ of
 the variations (\ref{vs3}) just cancels its inverse entering
 the inverted propagators. Hence, the BRS-variation 
 of the Gaussian measure has mass dimension $ D = 5$. Requiring the 
 regularized generating functional $Z^{\Lz}(K)$, (\ref{vs1}),
 to be invariant under the  BRS-variations (\ref{vs3})
of the integration variables,
 provides the \emph{violated Slavnov-Taylor identities} (VSTI)
 \begin{equation} \label{vs5}
    0 \stackrel{!}{=}
 \int d\mu_{\Lz}(\Phi) \  e^{-\frac{1}{\hbar} L^{\LLzz}(\Phi)
 +\frac{1}{\hbar} \langle \Phi , K \rangle }
 \Bigl( \delta _{BRS}\, \langle \Phi , K \rangle - \delta _{BRS}\,
        ( Q^{\Lz} + L^{\LLzz}) \Bigr)\ .
 \end{equation}  
 The BRS-variations appearing in (\ref{vs5}) can be  dealt with,
 considering corresponding modified generating functionals: \\
 i) With the modified bare interaction (\ref{i4}) we define
 \begin{equation} \label{vs6}
{\tilde Z}^{\Lz} (K,\xi ) := \int d\mu_{\Lz}(\Phi)
    \,e^{-\frac{1}{\hbar} {\tilde L}^{\LLzz}(\xi ; \Phi) 
+\frac{1}{\hbar} \langle \Phi , K \rangle }\ ,
 \end{equation} 
 and introduce a \emph{regularized}  BRS-operator 
 \begin{equation} \label{vs7}
 {\mathcal D}_{\Lambda_0} = \sum_{\tau} 
 \big \langle \Jt \, , \sio\, \frac{\delta }{\delta {\gamma}_{\tau}}
             \big\rangle
 + \big \langle \etb, \sio \frac{\delta }{\delta \omega ^{a}} \big\rangle  
 +\big\langle \, \frac{1}{\alpha }
 \,\partial _{\nu} \frac{\delta }{\delta \jn}
    - m \frac{\delta}{\delta \ba}\, ,\sio \et \big\rangle\  .
 \end{equation} 
ii) The BRS-variations of the bare action and of the Gaussian measure 
 \begin{equation} \label{vs8}
 L^{\LLzz}_1 \eps :\, = \,
      - \delta _{BRS} \Big( Q^{\Lz} + L^{\LLzz} \Bigr) 
         = \int dx\,  N(x) \,\eps 
 \end{equation} 
 form  a space-time integrated insertion with ghost number $1$.
 The variation of $ L^{\LLzz} $, however, keeps the regularizing 
 factor $ \sio$ of (\ref{vs3}), thus the integrand $ N(x) $ is no longer
 a polynomial in the fields and their derivatives. 
 We can initially treat the integrand $ N(x) $ as a local insertion
  with a source $\rho(x)$, cf. (\ref{i11}). Introducing the corresponding
 bare action  $ {\tilde L}^{\LLzz}(\rho ; \Phi) $ similarly to 
 (\ref{i4}), we define  the
 functional \footnote{ Abusing notation we let the variables $\rho$ and
 $\xi$, respectively, denote different functions.}  
${\tilde Z}^{\Lz} (K,\rho ) $ in analogy to (\ref{vs6}). \\ 
 In terms of these modified $Z$-functionals
 the VSTI
 (\ref{vs5}) can now be written 
 \begin{equation} \label{vs9}
  \mathcal{D}_{\Lambda_0} \, \tilde{Z}^{\Lz} (K, \xi ) |_{\,\xi =0} \, = \, 
   \int dx \frac{\delta }{\delta \varrho (x)}
 \tilde{Z}^{\Lz} (K, \rho ) |_{\,\rho =0}\ . 
 \end{equation} 
 The modified $Z$-functional (\ref{vs6}) is related to the corresponding
   generating functional of modified CAS by
   \footnote{ The vacuum part $ I^{\Lz} $ is the same as in the case
 without insertion, since the latter has nonzero ghost number} 
 \eq \label{vs10}
 {\tilde Z}^{\Lz} (K,\xi )  =  e^{\frac{1}{\hbar} P^{\Lz}(K) } \,
  e^{-\frac{1}{\hbar} ( {\tilde L}^{\Lz}
 ( \xi ; \,\fit , \, c, \,\bar c ) + I^{\Lz}) } \ ,
 \eqe
 and analogously in case of ${\tilde Z}^{\Lz} (K,\rho ) $.
 Furthermore, the variables of the $Z$- and the $L$-functional satisfy
\begin{equation}  \label{vs11}
\begin{split}
  \fitx &=  \int dy \, C^{\Lz}_{\tau} (x-y)\, \Jt (y)\ , \\
  \cax &=  - \int dy \,S^{\Lz} (x-y)\, \et (y)\ ,  \qquad
         \cbx  =  - \int dy\, S^{\Lz} (x-y)\, \etb (y)\ . 
 \end{split}        
\end{equation} 
 From (\ref{vs9}), via (\ref{vs10}) and the analogous relation
 for ${\tilde Z}^{\Lz} (K,\rho ) $, we derive,
using the definitions (\ref{i7}), (\ref{i12}) and denoting the
differential operators (\ref{f1}) by $D_{\tau} $ in accord with $\fit $,
 the \\ 
\emph{ violated Slavnov-Taylor identities of the CAS:} 
\begin{eqnarray} \label{vs12} 
 \big \langle \ca , D\, \big (\, \frac{1}{\alpha }\, \partial _{\nu}
 \An - m\B \big) \big \rangle -
 \big \langle \ca ,\sio \big(\, \partial _{\nu} 
  \frac{\delta L^{\Lz}}{\delta \An} - m\frac{\delta L^{\Lz}} {\delta \B}
       \big) \big\rangle \nonumber \\
 + \sum_{\tau} \big \langle \fit \, , D_{\tau} L^{\Lz}_{\gamma _{\tau}}
  \big \rangle - \big \langle \cb , D L^{\Lz}_{\oma} \big \rangle \, 
               = \, L^{\Lz}_1\ . \qquad \qquad
\end{eqnarray}

  Starting from the relations (\ref{vs21}) 
between the generating functionals of the vertex- and
Schwinger-functions we can convert  (\ref{vs12})
at the (physical) value $ \La = 0 $  into the 
\emph{violated Slavnov-Taylor identities for proper vertex functions},
on  substituting there the fields $ \Phi $ according due to (\ref{vs19}), and
  employing (\ref{vs20}), (\ref{vs21}), 
\begin{equation} \label{vs22}
\begin{split}
 \sum_{\tau} \Big\langle \frac{\delta \Gamma^{\Lz} }{\delta \ufit} 
  \, ,\sio \Gamma^{\Lz} _{\gt} \Big\rangle 
           - \Big\langle \frac{\delta \Gamma^{\Lz}}{\delta \uca} ,\,\sio
             \Gamma^{\Lz} _{\oma} \Big\rangle  
  - \Big\langle \frac{1}{\alpha }\, \partial _{\nu} \uAn - m\uBa,\,
              \sio \frac{\delta \Gamma^{\Lz} }{\delta \ucb}\Big\rangle  \\
              = \, \Gamma^{\Lz} _1 ( \ufit \, , \uca, \ucb ) \ ,
         \qquad \qquad \qquad \qquad \qquad
 \end{split}             
 \end{equation} 
 with
 \begin{equation} \label{vs23}
   \Gamma^{\Lz} _1 (\ufit , \uca, \ucb ) \, = \, 
    L^{\Lz}_1 ( \fit , \ca, \cb ) \ .
 \end{equation}

In the analysis of the STI it will turn out that we need the form of 
their explicit violation ``on the bare side'', $\,\Ga^{\LLzz}_1(\uFi) \,$,
 too.
From the definition (\ref{vs8}) we directly determine the bare functional
 $ \,L^{\LLzz}_1(\Phi) \,$, using (\ref{i4}) and (\ref{i5}),
\begin{equation} \label{stiv1} 
  L^{\LLzz}_1(\Phi) \,=\, 
  \langle \ca , D\, (\,\frac{1}{\alpha }\, \partial _{\nu} \An - m\B) \rangle
  + \sum_{\tau} \langle \fit \, ,D_{\tau} L^{\LLzz}_{\gamma _{\tau}} \rangle 
  - \langle \cb , D L^{\LLzz}_{\om}\rangle 
\end{equation} 
  $$
  - \,\Big \langle \frac{\delta L^{\LLzz}}{\delta \cb}, \si_{\Lz} 
  (\, \frac{1}{\alpha }\, \partial_{\nu} \An - m\B) \Big \rangle  
 + \sum_{\tau} \Big \langle \frac{\delta L^{\LLzz}}{\delta \fit} , \si_{\Lz}
  L^{\LLzz}_{\gamma _{\tau}} \Big\rangle 
 - \Big \langle \frac{\delta L^{\LLzz}}{\delta \ca}, 
      \si_{\Lz} L^{\LLzz}_{\om} \Big\rangle .
 $$
The functional $\,  L^{\LLzz}_1(\Phi) \,$
generates $n$-point functions with $\, 2 \leq |n| \leq 5$. Moreover, we
 observe that only the terms emerging from the BRS-variation of the
 bare interaction $\, L^{\LLzz}\,$ have mass dimension  greater than
  $\, D=5 \,$, because of the cutoff function $\, \si_{\Lz}(k^2) \,$
 (cf. remark after (\ref{vs4})). Given the functional $\, L^{\LLzz}_1 \,$,
 its $n$-point functions coincide with those of the functional
   $\, \Ga^{\LLzz}_1 \,$, due to the identity (\ref{clv3}).\\
   
\section{Restoration of the Slavnov-Taylor Identities}
\subsection{Mass expansions of Vertex Functionals}
To restore the STI, it is in particular  necessary
to make vanish the relevant part of the violating functional
$\, \Gamma^{\,0,\,\Lambda_0}_1\, $. It will then turn out that 
this is also sufficient in the limit $\Lao \to \infty\,$.
Namely the irrelevant contributions to this functional 
at the bare scale  $\, \Gamma^{\,\Lambda_0,\,\Lambda_0}_1\, $,
which stem from the regulating function $\sigma_{0,\Lambda_0}\,$, are 
sufficiently bounded in terms of inverse powers of $\Lao\,$
so that we may apply Proposition 3 providing the
bound (\ref{bdf}). 

The freedom we dispose of to achieve this task is the freedom of
choosing the renormalization conditions for  the relevant terms
appearing in the functionals $\, \Gamma^{\,0,\,\Lambda_0}_{l,n}\,$
and $\, \Gamma^{\,0,\,\Lambda_0}_{\ga ;\, l,n}\, $.
On inspection of the VSTI (\ref{vs22}) one realizes 
that there is an obstacle on this way of proceeding~:
Since the insertion defining the functional  
$\, \Gamma^{\,\Lambda,\,\Lambda_0}_1\, $ is of dimension 5,
we have to apply up to 5 field- and momentum-derivatives 
on  (\ref{vs22}) in order to exhaust all relevant terms. 
We first notice that momentum derivatives of the cutoff function
$ \, \si_{\Lz}( k^2) = \si_{\Lao}(k^2)  \,$ do not contribute
to the relevant terms looked for,
\footnote{ This property  is at the origin of our
particular choice of the cutoff function.} cf. (\ref{f6}). Hence,  
in the terms generated from (\ref{vs22}) by these field- or 
momentum-derivatives
there apply $d_1$ (field or momentum)-derivatives 
to the factors of the form $\de \Ga/ \de \vp\,$  in (\ref{vs22}), 
and  $d_2\,$ (field or momentum)-derivatives apply to the factors 
of the form  $\,\Ga_{\ga}\,$,  $\pa A^a\,\,$, or $m B^a\,$, 
where $\,d_1+d_2\,\le 5$. If $d_2 \ge 3\,$
derivatives apply to the functionals 
$\, \Gamma^{\,0,\,\Lambda_0}_{\ga ;\,l,n}\, $, they 
generate irrelevant contributions, since 
the insertions in $\, \Gamma^{\,0,\,\Lambda_0}_{\ga ;\,l,n}\, $ are of
dimension 2. In our earlier paper [KM]
such contributions to the VSTI were denoted by "irr"
in its Appendix C. They hampered
the analysis of the relevant part of the VSTI at the renormalization 
scale in our previous
efforts since they cannot be controlled  {\it explicitly} in terms 
of the renormalization conditions.
The only way out can be that the relevant terms
from $\, \Gamma^{\,0,\,\Lambda_0}_{l,n}\, $ multiplying these irrelevant   
terms can always be made to vanish so as to avoid the a priori
unknown irrelevant terms to appear. One then realizes however that there
are contributions in   $\, \Gamma^{\,0,\,\Lambda_0}_{l,n}\, $, present
already at the tree level $l=0\,$, which do not satisfy this criterion, 
namely the nonvanishing super-renormalizable three-point couplings,
as well as the mass term of the $2$-point functions 
(see Appendix A).

We present the following solution to this problem :  The functionals 
$\, \Gamma^{\,0,\,\Lambda_0}_{\ga ;\,l,n}\,$ and 
$\, \Gamma^{\,0,\,\Lambda_0}_{l,n}\, $ are expanded at zero momentum
not only w.r.t. the fields and the momenta but also w.r.t. to
 the number of super-renormalizable vertices, or otherwise stated
w.r.t. to the number of mass parameters appearing in these couplings,
see Section 3.2, (\ref{nv1}) and (\ref{nv2}). 
The degree of divergence then diminishes with this number,
in fact the corresponding  bounds (\ref{bv1}) and (\ref{bv2})
show that the presence of an explicit mass term produces 
a gain in power counting by one unit.
Disposing then of all relevant terms in this new sense, 
we will realize that there do not remain 
uncontrollable contributions to the VSTI of the form mentioned above.
One should note that counting a power of
a mass parameter as a power of a field, is intuitively 
in accord with the fact that
these mass parameters stem from the vacuum expectation value of the
scalar field.\\
We start introducing the expansion of the
functionals $ L^{\LLz}_1$  and $\Ga^{\LLz}_1 $ inherited from 
the mass scaling (\ref{scale}),
 \eq
{\mathcal L}^{\La,\Lao}_{1;\,l, n}(\la;\vec{p}\,) \,=\,
\sum_{\nu=0}^{\infty}(m \la)^{\nu}\
{\mathcal L}^{(\nu),\La,\Lao}_{1;\,l, n}(\vec{p}\,) \ ,\quad
 \vec{p}= (p_1,\cdots,p_{|n|})\, ,
\label{ma1}
\eqe
 \eq
{\Ga}^{\La,\Lao}_{1;\,l, n}(\la;\vec{p}\,) \,=\,
\sum_{\nu=0}^{\infty}(m \la)^{\nu}\
{\Ga}^{(\nu),\La,\Lao}_{1;\,l, n}(\vec{p}\,) \ .
\label{ma2}
\eqe
Since we aim at a consistent mass expansion of the VSTI, 
(\ref{vs22}), we first observe, that we also have to perform the mass
 scaling (\ref{scale}) of the BRS-variation 
$\,\frac{1}{\alpha} (\partial_{\nu}A^a_{\nu}(x) - \alpha m B^a(x)) \,$
 of the antighost appearing, cf.(\ref{y13}), in accord 
with our treatment of the BRS-insertions.
We then want to determine via (\ref{vs22}) the relevant part
 of the functional $\, \Ga^{\Lz}_1 $, given by the values 
 $\,( \pa^{w} \Ga^{(\nu),\, 0, \,\Lao}_{1;\,l, n}) (\vec 0) \, ,
 \,\, |n|+|w|+\nu \leq 5 \,$.
It is important  to note that  irrelevant  contributions only emerge
from the functionals containing a BRS-insertion.
 Requiring the vertex functions in (\ref{vs22}) to satisfy 
 the boundary conditions, $\, l \in {\mathbf N}_0 $,
 \eq \label{ma3}
 ( \pa^{w} \Ga^{(\nu),\, 0, \,\Lao}_{ l,\, n}) (\vec 0) \,
  \stackrel{!}{=}\, 0 \,, \quad
  \mbox{if} \quad |n|+|w|+\nu <  4 \,,
 \eqe
 irrelevant contributions  
 from the functionals $ {\Ga }^{(\nu), \Lz}_{\ga_{\tau}} ,
 {\Ga }^{(\nu), \Lz}_{\om} $  then are annihilated by multiplication 
 and only contributions of these functionals with 
 $ |n_2|+|w_2|+ \nu_2 \leq 2\,$ field-, momentum- and 
mass-derivatives, i.e. relevant terms, do  appear. 
The  condition (\ref{ma3}) is
 satisfied for $ \, l\geq 1 \,$ by the renormalization
 conditions (\ref{nv5}), and in the tree order, 
if $ |n|=3\,$, (\ref{nv3}). \\
Here, we remind the reader that we do not apply the mass
expansion to the free propagator, but only to the boundary terms
appearing in the FE. Now the inverted free propagators
$\Ga ^{0,\Lao}_{0,\,n},\, |n| = 2 ,\,$ appear in (\ref{vs22}) 
as boundary terms at $\La=0\,$ for the functions 
$\Ga ^{\La,\Lao}_{1;l,n}\,$, and they are then mass expanded,
(\ref{scale}), thus satisfying (\ref{ma3}), too.    
Therefore it is important to remember that the FE and the VSTI
are derived {\it before} mass expanding. 
Afterwards we consistently apply the mass expansion to all boundary
terms and make the corresponding statement on the bounds for the 
vertex functions which is verified inductively.
  
 The renormalization conditions (\ref{nv5}) imposed
 on (a subset of) the relevant 
 terms of the vertex functions imply
 zero renormalization conditions for the leading 
contributions to all the two-point functions~:
\eq
\de m^2_{(\nu)} = 0\ ,\quad
 \Sigma^{\bar{c}c(\nu)}(0)=0\ ,\quad \Sigma^{BB(\nu)}(0)=0\ ,\quad 
 \Sigma^{hh(\nu)}(0)=0\  \mbox{ for }\ \nu \le 1\ ,
\label{2pt}
\eqe
and also
\eq
\Sigma^{AB(\nu)}(0)=0  \    \mbox{ for }\  \nu = 0~;\quad
\ka^{(\nu)} =0\    \mbox{ for }\  \nu \le 2\ .
\label{21pt}
\eqe
Here we use the notations of App. A. The respective relevant parts of
the inserted functionals $\Ga_{\ga}\,$ are collected in App. B.
The restricted set of renormalization conditions (\ref{nv6}) is 
automatically satisfied, even in the nonvoid case with $ |n|=1 $,
\eq \label{ma4}
  n \equiv c^a \,:  \quad 
 \Ga^{\Lz}_{\ga^a ;\, n}(0; \vec 0\,) = m\, R_4\, ,
\eqe
due to the explicit factor of $m$ to be scaled
according to (\ref{scale}).

The functionals  $\, L^{\LLz}_1(\Phi), \,\Ga^{\LLz}_1(\uFi) \,$ serve to
 control the violation of the STI. They contain irrelevant boundary terms
 at $\La =\Lao\,$, in contrast to the functionals without
 insertion or with a BRS-insertion. These boundary terms are due 
to the presence of the factors
$\,\si_{0,\Lao} $, cf. the remarks 
after (\ref{stiv1}). They are proportional to
 $\,\si_{0,\Lao}(p)-1 =O((p^2)^2/\Lao ^4)\,$, 
  as follows from (\ref{f5}),  
since the terms proportional to $\si_{0,\Lao}(0)=1\,$
are relevant. \\
  We first assert the  bound on the bare functional
  $ \Ga^{\LLzz}_1 $, valid for $ l \in {\mathbf N}_0 $,
 \eq \label{ma5}
 |\,\partial^w \Ga^{(\nu),\, \LLzz}_{1;\,l, n}( \vec p \,)| \,\leq
  \,(\Lao +m)^{\,5 -|n|-|w|- \nu} \,
   \Big ( \log{\Lao\over m}\Big)^{r} \,
    {\cal P}(\frac{|\vec p \,|}{\Lao})\ ,
\eqe
 and trivially satisfied, unless $ 2 \leq |n| \leq 5 $. Because of the 
 identity (\ref{clv3}) we can establish the corresponding bound on 
 $ L^{\LLzz}_1 $ and  making use of (\ref{stiv1}). We employ 
 the previous bounds on $ \pa^{\, w} {\mathcal L}^{(\nu), \,\LLz}_{l, n} $,
 (\ref{b1}), and on
 $ \pa^{\, w} {\mathcal L}^{(\nu),\, \LLz}_{1;\,l, n} \,$,
 (\ref{b2}) , at the value $ \La = \Lao $. For
  $ \si_{\Lz}(k^2) = \si_{\Lao}(k^2) $ we use the bounds
  $$
  |\, \pa^{\, w} \si_{\Lao}(k^2) | \,\leq \,
    \Lao^{ - |w|} \,{\mathcal P}_{|w|}\big ( \frac{|k|}{\Lao} \big ) \
    , $$
  which are an easy consequence of (\ref{f5}), the polynomials
 $ {\mathcal P}_{|w|} $ having nonnegative coefficients not depending on
 $ k $. With these ingredients we prove (\ref{ma5}).\\
The bound on the functional  $ \Ga^{\LLz}_1 $ (\ref{bdf}) 
 does not follow from the choice of standard renormalization
 conditions for insertions.
  We rather  assume its relevant part at the physical
 value $ \La = 0 $ of the flow parameter to vanish, 
 $\, l \in {\mathbf N}_0 \,$ ,
  \eq \label{ma6}
 ( \pa^{\, w} \Ga^{(\nu),\, \Lz}_{1; \,l, n})( \vec{0}\, )  = 0\, ,
      \qquad |n | + | w | + \nu \leq 5 \, .
 \eqe 
 In Section 5.3 we will be able to verify these conditions from 
 the VSTI (\ref{vs22}), choosing for the functionals
 entering the l.h.s. suitable renormalization conditions within
 the class (\ref{nv5}), (\ref{nv6})
  considered. Assuming (\ref{ma6}), we want to show 
that the corresponding irrelevant part vanishes upon
 shifting the UV- cutoff to infinity:  \\
 \textbf{Proposition 3}  \\
 \emph{ Given (\ref{ma6}), then for $ l \in {\mathbf N}_0 , |n| \geq 2 $
  and $ 0 \leq \La \leq \Lao $ ,}
 \eq \label{bdf}
 |\,\partial^w \Ga^{(\nu),\La,\Lao}_{1;\,l, n}( \vec p\, )| \,\leq
  \, \frac{1}{\Lao} \,(\La+m)^{\,5+1-|n|-|w|- \nu}
   \Big ( \log{\Lao\over m}\Big)^{r} \,
    {\cal P}(\frac{|\vec p\, |}{\La+m})\ .
\eqe
\emph{ with a positive integer $r$ depending on $ n, l, w\, $, and a
  polynomial ${\mathcal P} $ as in (\ref{b1}),(\ref{b2}). } \\
 \emph{Proof}: We first notice, that the bound (\ref{bdf}) at
 $\, \La = \Lao \,$ agrees with the bound (\ref{ma5}) , and at
 $\, \La < \Lao \,$ majorizes this bound, if  $ \, |n| + |w| + \nu >  5 $.
 The functions $ \,\partial^w \Ga^{(\nu),\La,\Lao}_{1;\,l, n} $
 with flow parameter $ 0 \leq \La \leq \Lao $  are bounded
 integrating inductively the FE  (\ref{fegi}), adapted
 to an integrated insertion and to the
  $ \lambda$-expansion, however, as stated.
We proceed in the inductive order as in the proof of the
  Proposition 1, but observing that the relevant terms of the
 functional treated here satisfy $\,|n| + |w| + \nu \leq  5 $.\\
 Considering the tree order first we notice, that the
 r.h.s. of the FE does vanish. Hence, this order is already fixed
 by its boundary value at $ \La = \Lao $. If $|n| = 2 $,
 the boundary value even vanishes and thus the function itself,
 satisfying (\ref{bdf}) trivially. Proceeding, for given $n$ 
 in the irrelevant cases $ |n|+|w| +\nu > 5 $ 
 the bound (\ref{bdf}) follows from the bound on their
 boundary values. Integrating the relevant cases 
 $ \,|n|+|w|+\nu \leq 5 \,$ with initial values (\ref{ma6}) yields
 $\, \pa^{\,w}\,\Ga^{(\nu), \LLz}_{1 ;\,0, n}({\vec 0}\,) = 0\, $.
Descending in $ |w| $, the integrand in the respective remainder
of the Taylor extension has already been bounded before,
providing the bound for general value $ {\vec p} \,$.
 Hence, the assertion is established in the tree order.\\
Proceeding for $\,l > 0\,$ inductively as indicated, the
$L$-functions appearing on the r.h.s. of the FE (\ref{fegi}) 
have to be determined within this inductive process
 via (\ref{clvi}), as expounded in presenting the FE 
 and supplemented in the text after (\ref{clv5}),
leading to the Proposition 2.
Therefore, to bound the r.h.s. one also needs the
bound (\ref{bv1}) on the vertex functions without
insertions, to be dealt with independently before.
As a result the bound deduced on
$\,| \pa^w L^{(\nu),\,\LLz}_{1 ;\, l-1,\,n'}| \,$  
essentially coincides with the bound on
$\,| \pa^w {\Ga}^{(\nu),\,\LLz}_{1 ;\, l-1,\,n'}| \,$,
 cf. (\ref{clv5}), i.e. has the same form and power
behaviour of $ \La+m $. This bound allows to estimate 
the r.h.s. of the FE and hereafter the integrations
"downwards" with initial conditions (\ref{ma5}),
and "upwards" with initial conditions (\ref{ma6}),
 of the irrelevant and relevant cases, respectively.
Extending finally the relevant cases via the Taylor
formula to general $\,\vec p\, $ completes the proof.
\qed
Thus, given the condition (\ref{ma6}),
the bound (\ref{bdf}) implies that  
 {\it the  Slavnov-Taylor-Identities are restored
in the limit  $\Lao \to \infty\,$.}

\subsection{Equation of motion of the anti-ghost}
Renormalization theory for nonabelian gauge theories in
gauge invariant renormalization schemes is generally  
based on the STI, complemented by the equation of motion
of the antighost [Z], [FS]. In our scheme we rather start 
from a derivation
of this equation from the functional integral.
In Section 5.3 we will then show that this equation
is satisfied for renormalization conditions compatible
with the STI if {\it in addition the renormalization
condition for the longitudinal part of the gauge field
propagator is fixed uniquely} to vanish at zero momentum. 

The field equation follows from the  
 representation (\ref{f14}). After functional derivation
 of  (\ref{f14}) with respect to
 $ \cbx $ we reexpress the r.h.s. as
 $$ \frac{\de L^{\LLz}(\Phi)}{\de \cbx}\,
    e^{-\frac{1}{\hbar}\left( L^{\Lambda ,\Lambda_0}(\Phi) 
               + I^{\Lambda ,\Lambda _0} \right)}
  = \frac{\de}{\de \zeta^{a}(x)} 
   \int d\mu _{\Lambda ,\Lambda _0}(\Phi ' )\ e^{-\frac{1}{\hbar}
               \big ( L^{\Lambda_0 ,\Lambda _0}( \Phi ' +\Phi ) 
       + L^{\LLzz}(\zeta;\, \Phi' + \Phi) \big ) } {\Big|}_{\zeta = 0}  
     $$
  on extending  the original bare interaction $ L^{\LLzz}(\Phi) $
  by the insertion
  \eq \label{vs14}
      L^{\LLzz} (\zeta ; \Phi) \, = 
          \int dx \ \zeta^a (x)\  \frac{\de L^{\LLzz} (\Phi)}{\de \cbx}\ .
 \eqe
The source $ \zeta^a(x) $ is a Grassmann element carrying
 ghost number $ -1$.  Treating now the r.h.s. analogously as in
 (\ref{i4}) - (\ref{i7}),
 we obtain the field  equation of the antighost     
\eq \label{vs13}
 \frac{\de L^{\LLz} (\Phi)}{\de \cbx} \,=\, L^{\LLz}_{\zeta^a }( x;
 \Phi)\ ,
 \eqe
 employing the notation introduced there.
 On the r.h.s. appears the generating functional of the CAS 
   with one local insertion corresponding to (\ref{vs14}). 
  The classical BRS-invariant
   action (\ref{y9a}) satisfies the classical field equation
   $ \de / \de \cbx S_{BRS} = 
    \partial_{\mu}\psi^a_{\mu}(x) - \alpha m \psi^a (x) $,
   observing (\ref{y14}). 
The aim is to show that the relation following from the classical
action at the tree level for the 
physical value $ \La = 0 $ of the flow parameter
  \eq \label{vs15}
   \frac{\de L^{\Lz} (\Phi)}{\de \cbx} \,=\,
    \pa_{\mu} L^{\Lz}_{\g_{\mu}}(x\, ; \Phi) |_{mod}
     - \alpha m L^{\Lz}_{\g}(x ; \Phi) |_{mod} \,,
    \eqe
still holds
in the  renormalized theory. The label
 "mod" is to signal that we have to replace in the
 bare insertions (\ref{3.12a})-(\ref{3.12d}) 
  $ R^0_i \rightarrow { \tilde R}^0_i = O(\hbar) $
 for $ i=1,4 \,$  since the respective tree order 
is absent on the l.h.s.

We can write (\ref{vs15})
in terms of proper vertex functions. Fourier transforming
 (\ref{vs15}), using our conventions (\ref{f21}), (\ref{i8}),
 and employing the relations (\ref{vs21}), (\ref{vs20}) yields
 \eq \label{vs28}
   (2\pi)^4 \, \frac{\de {\Ga}^{\Lz} (\uFi)}{\de \ucb(q)} \,=\,
    - \, \frac{q^2 + \al m^2}{\sio(q^2)}\, \uca(-q) 
    - \, i q_{\mu} {\Ga}^{\Lz}_{\g_{\mu}}(q ; \uFi) |_{mod}
     -\, \alpha m {\Ga}^{\Lz}_{\g}(q ; \uFi) |_{mod} \,.
    \eqe 
  The first term on the r.h.s. is the tree level
 $2$-point function. Restricting
  (\ref{vs28}) to its relevant part, $\sio(q^2) $  is
  replaced by $ \sio (0) = 1 $ due to (\ref{f6}), 
the first term then provides the 
  tree order of $R_1$ and $R_4$ excluded in the insertions as
  indicated by the  label \emph{mod}, cf. (\ref{vs15}). 

The proof of (\ref{vs28}) or equivalently (\ref{vs15})
consists in two steps of the same nature as those employed
in the previous section.
We may consider the (regularized) inserted functional
\eq
 \label{gh1}
\Ga_{c^a}^{\La,\Lao}(q\,;\uFi)~:=
    (2\pi)^4 \, \frac{\de {\Ga}^{\La,\Lao} (\uFi)}{\de \ucb(q)} \,+\,
     \frac{q^2 + \al m^2}{\sio(q^2)}\, \uca(-q) 
    + \, i q_{\mu} {\Ga}^{\La,\Lao}_{\g_{\mu}}(q ; \uFi) |_{mod}
     +\, \alpha m {\Ga}^{\La,\Lao}_{\g}(q\,; \uFi) |_{mod} \ .
    \eqe 
In the mass expansion scheme it corresponds to an operator 
insertion of dimension 3, where we take into account also
the momentum and mass factors in front of the last three terms. 
Since the flow equations for inserted functionals are linear,
the new functional obeys again a linear flow equation
obtained from those for the functionals on the r.h.s.
by superposition. Note that the second term on the r.h.s., 
being a tree level contribution,
does not flow.\\ 
If we can choose renormalization conditions such that
all relevant contributions to 
$\Ga_{c^a}^{\La,\Lao}(q\,;{\uFi})\,$ vanish, we can 
prove by induction on the linear flow equation
(the solution of which is unique for specified boundary conditions)
 that $\Ga_{c^a}^{\La,\Lao}(q\,;\uFi) \equiv 0\,$.
Note that for this functional there are no irrelevant boundary
contributions at $\La=\Lao\,$,
since such terms  only  appear in the first two terms
on the r.h.s. at the tree level and cancel exactly.
So the situation is simpler than that of the functional
$\Ga_1\,$ analysed in the previous section.

At the end of the next section it is shown explicitly that the relevant
contributions to (\ref{gh1}) can be made to vanish for suitable 
renormalization conditions so that 
{\it the equation of motion for the antighost  (\ref{vs28}) or (\ref{vs15})  
 holds at the quantum level}.

\subsection{Analysis of the relevant part of the  Slavnov-Taylor
  Identities and of the equation for the antighost \label{relsti}}
 We now require the relevant part of the functional
 $\, \Gamma^{\,0,\,\Lambda_0}_1 $ to vanish in accord
 with the VSTI (\ref{vs22}).
 This requirement amounts to satisfy the $53$ equations
 presented in the Appendix C.
 It is satisfied in the tree order.
Noticing that the normalization constants of the BRS-insertions behave as  
$\, R_i = 1+ {\mathcal O}(\hbar) , i = 1,\cdots 7$, we first analyse the
 equations $ IX $ to $ XXIX $, but take already into account
the equations $ VII_d \,, \, VIII_c $ \,, the latter ones providing
\begin{equation} \label {c1}
    r^{hBA}_2 \, = \, r^{\bar{c} c A}_2 \, \stackrel{!}{=} 0 \, .
\end{equation} 
In proceeding we use conditions determined before, if needed. \\ 
From $XIV_b \,, \,XIV_e \,, XV_{1b}\,, XXIII \, $ directly follow
 \begin{equation}\label{c2}
  r^{AA{\bar c}c}_1 \, = \, r^{AA\bar{c} c }_2 \, =   
  r^{BB{\bar c}c}_1 \, = \, r^{AABB}_2 \, \stackrel{!}{=} 0 \, ,
 \end{equation} 
and than, from $ XIV_{a+c}\,, XVII_b \,, XVIII_c \,, XXVIII, XXIX $ ,
 \begin{equation} \label{c3}
  r ^{AAAA}_2 \, = \,r^{hh{\bar c}c} \, = \, r^{{\bar c}c \bar{c}c} \, =   
  r^{hB{\bar c}c} \,= \,   r^{BB{\bar c}c}_2\,\stackrel{!}{=} 0 \, .
 \end{equation}     
 $ XVI_a \,, XVIII_a \,$, and $XV_{2a}$ combined with $\,XVI_b \,$,
 respectively, require
\begin{equation}\label{c4}
 R_2 \, \stackrel{!}{=} \,R_6 \, \stackrel{!}{=} \,R_7 \,, \qquad \quad
   R_3 \, R_5 \, \stackrel{!}{=} \,( R_2 )^2 \,.
   \end{equation}
   \begin{equation}\label{c5}
     XIV_c : \qquad  2 F^{AAAA}_1 \, R_1 \,
            \stackrel{!}{=}\,- F^{AAA} g R_2 \,
   \end{equation}  
   \begin{equation}\label{c6}
     XI  : \qquad \quad F^{{\bar c}c B\, (1)} \, R_5 \,
                    \stackrel{!}{=}\,- F^{{\bar c}c h \,(1)} R_2 \,.
   \end{equation} 
 From $ X \,, \,XX\,, \, XIX\, , \, IX \,$ follow for the
                self-coupling of the scalar field 
 \begin{eqnarray}
    8\, F^{BBBB} \, R_4 & \stackrel{!}{=} & F^{BBh (1)} \,g R_3 \,,
                           \label{c7} \\
     4\, F^{BBhh} \, R_4 & \stackrel{!}{=} & F^{BBh (1)}\, g R_5\,,
                              \label{c8} \\
     8\, F^{hhhh} \, R_4 R_3 & \stackrel{!}{=} & F^{BBh (1)}\, g (R_5 )^2 \,,
                               \label{c9} \\
     F^{hhh \,(1)} \, R_3 & \stackrel{!}{=} & F^{BBh (1)} \,R_5  \,,
                                 \label{c10}
     \end{eqnarray}
  and from $XVI_b\, ,\, XVII_a \,,\, XXI\,,\, XIII_2 \,$ for the
                         scalar-vector coupling
 \begin{eqnarray}
 2\, F^{BBA} \,R_5  & \stackrel{!}{=} & - \,F^{hBA}_1\, R_2 \,,
                            \label{c11} \\
 4\, F^{AAhh} \,R_1  & \stackrel{!}{=} &  \,F^{hBA}_1 \,g R_5 \,,
                                \label {c12} \\
 4\, F^{AABB}_1 \,R_1  & \stackrel{!}{=} &  \,F^{hBA}_1\, g R_3 \,,
                                 \label{c13}\\
      F^{AAh (1)} \,R_1  & \stackrel{!}{=} &  \,F^{hBA}_1\,  R_4 \,.
                                 \label{c14} 
 \end{eqnarray}     
One easily verifies that the remaining equations of $ IX $ to $ XXIX $
   are satisfied due
to these conditions (\ref{c1})-(\ref{c14}).\\
At this stage, all those relevant couplings with $\, |n| = 3,4\,$ not 
  appearing already
 in the tree order are required to vanish: (\ref{c1})-(\ref{c3}).
   All other couplings
 involving four fields are determined by particular couplings
  with $ |n|=3\, $: (\ref{c5}),
 (\ref{c7})-(\ref{c9}), (\ref{c12}),(\ref{c13}). In addition, there
  are $4$ conditions
 relating couplings with $ |n|=3 $ : (\ref{c6}), (\ref{c10}), (\ref{c11})
       and (\ref{c14}).
 Moreover, the normalization constants of the BRS-insertions 
 are required to satisfy the three conditions (\ref{c4}).\\
 \noindent
 There are still $ 18-2 $ equations among $ I $ to $ VIII $ to be considered.
 They contain the relevant parameters of $\,\Gamma^{\,0, \Lambda_0}\,$
 with $ |\,n| = 1,2,3 \,$, except $ F^{hhh} $,
 together with the normalization constants of the BRS-insertions.
 Since $2$ of these parameters have been fixed before, (\ref{c1}),
  there remain
  $26 $ to be dealt with.\,($ F^{hhh} $ will then be determined
  by (\ref{c10}).)
These parameters in addition have to obey the conditions derived before: 
 We first observe that the condition (\ref{c14}) is identical to
 equation $ VI_b \, $. 
 There remain the $5$ conditions  to be satisfied:
$3$ conditions (\ref{c4}), together
 with (\ref{c6}), (\ref{c11}). 
All these conditions generate $4$ linear relations among the 
  equations still to be considered: denoting by $ \{X\} $ the content
 of the bracket
 $ \{ \cdots \} $ appearing in equation $ X $, we find [M,\,(4.94-97)]   
 \begin{eqnarray}
 0 &=& \alpha^{-1} \{VIII_b \} +gR_2 \{I_b\} + 
        R_1 \big( \{III_a\} + \{III_b\}\big )\, , \label{c15} \\
 0 &=& gR_2 \{II_b\} - \{ VIII_b\} + R_1 \{IV_b\} - 2 R_4 \{V\} \, ,
  \label{c16}\\
 0 &=& R_2 \{ IV_a\} - R_3 \big ( \{VI_a\} - \{ VI_b\} \big ) \, ,
 \label{c17}\\
 0 &=& R_2 \{V \} - R_3 \{ VII_c \} \, .\label{c18}
 \end{eqnarray} 
 Hence, the $ 26 $ parameters in question are constrained by
 $ 16+5-4 = 17 $ equations.
 As \emph{renormalization conditions} we then fix
 $ \kappa^{(3)} = 0 \,$ and let
 \begin{equation}\label{c19}
 \Sigma_{\,\rm trans}\, , \Sigma_{\,\rm long}\, , \Sigma^{AB(1)} ,
 {\dot \Sigma}^{{\bar c} c},
    {\dot \Sigma}^{BB} , F^{AAA} , F^{BBh(1)} , R_3
 \end{equation}
 be chosen freely. These parameters correspond  to the number
of wave function renormalizations (including one for the BRS sector)
and coupling constant renormalizations of the theory.  
Thus, there are $ 26 - 9 $ parameters left,
 together with $ 17 $
 equations. These parameters are now determined successively in
 terms of (\ref{c19})
 and possibly parameters determined before in proceeding. We list
 them in this order,
 writing in bracket the particular equation fulfilled:
$$ R_1(I_b)\, , R_4(II_b)\,, R_2(III_b) \rightarrow R_6 , R_7 , R_5 
 \quad { \rm due \,to}\quad (\ref{c4}) \,, $$
  $$\,F^{{\bar c} c A}_1(III_a)\, , 
  F^{BBA} (V) \rightarrow  F^{hBA}_1 \quad {\rm due \,to}\quad
 (\ref{c11}) \, , $$
$$  F^{AAh(1)}( VI_b ) \, , \,F^{{\bar c} c B(1)}(IV_a) \rightarrow
    F^{{\bar c} c h(1)}\quad {\rm due \, to} \quad (\ref{c6}) \, ,$$
  \begin{equation} \label{c20}
  \Sigma^{{\bar c} c (2) }(VIII_a)\,, \,\Sigma^{BB(2)}(II_a) \, , \,
 \de m^2_{\,(2)}(I_a)
   \, , \, \Sigma^{hh(2)}(VII_a) \, ,\, {\dot \Sigma}^{hh}(VII_{b+c}) \,.
\end{equation}
Now all parameters are determined, without using the
 equations $ IV_b \,, 
  VI_a\, , VII_c \, , VIII_b \,.$ These equations, however,
 are satisfied because of 
 the relations (\ref{c15})-(\ref{c18}). Finally, the relevant
 couplings with $ |n| = 4 $,
 as well as $ F^{hhh(1)} $ , then are explicitely given
 by (\ref{c5}), (\ref{c7})-(\ref{c10}),
 (\ref{c12}) and (\ref{c13}). 
 
 We have not yet implemented the field equation of
 the antighost (\ref{vs28}).
Performing the mass scaling as before and then extracting the local content 
 $ |n|+ |w| + \nu  \leq 4 \,$ leads to the relations   
\begin{eqnarray}
 1 + {\dot \Sigma}^{{\bar c} c} & = & R_1 \, , \label{c22} \\
     \alpha + \Sigma^{{\bar c} c (2)} & = & \alpha R_4 \, , \label{c23} \\
            F^{{\bar c} c A}_1 & = & g R_2 \, , \label{c24} \\
   F^{{\bar c} c B(1)}\ & = & \frac{\alpha}{2} \,g R_6 \,, \label{c25}\\
  F^{{\bar c} c h(1)}\,  & = & - \,\frac{\alpha}{2}\, g R_5 \,. \label{c26} 
  \end{eqnarray}
   Fixing now the hitherto free renormalization constant
 $ \Sigma_{\rm long} \,$ at 
   the particular value  $ \Sigma_{\rm long} = 0 \,$, we claim these
 relations to be satisfied: 
   (\ref{c22}) and (\ref{c24}) follow at once from $ I_b $
   and $ III_{a+b} $, respectively;
   (\ref{c25}) follows from $ 2\{IV_a \} - \{IV_b \} $, due
 to (\ref{c24}) and (\ref{c4}); and
   herefrom follow (\ref{c26}) due to (\ref{c6}), and (\ref{c23})
 because of $ VIII_a \,$, thus establishing the claim. \\
 Given these additional relations (\ref{c22})-(\ref{c26}) we can adjust
 the procedure (\ref{c20}) choosing now a \emph{reduced set of free
  renormalization conditions (\ref{c19}) in which} $\Sigma_{\rm long}$
  \emph{is excluded.}
 Proceeding similarly as before we find
  \begin{equation} \label{c27}
  I_b : \quad \Sigma_{\rm long} \, = \, 0 \,,\qquad
         II_a : \quad \Sigma^{BB(2)} \,= \, 0 \, ,
  \end{equation}  
  \begin{equation} \label{c28}
     III_b  : \qquad g R_2 \, = \,  - \, 2 F^{AAA}\,
        \frac{ 1 + {\dot \Sigma}^{{\bar c} c}}{ 1+ \Sigma_{\rm trans}}  
 \longrightarrow \, R_6,\, R_7,\, R_5 \quad {\rm due \, to} \quad (\ref{c4}),
 \end{equation}
 \begin{equation} \label{c29}
  II_b  : \qquad   R_4 \,= \,   
          \frac{ 1 + {\dot \Sigma}^{{\bar c} c}}{ 1 + {\dot \Sigma}^{BB}}
                        \Big ( 1+ \Sigma^{AB(1)} \Big) \, ,
  \end{equation}   
  \begin{equation} \label{c30}
  I_a  : \qquad   1+ \de m^2_{\,(2)} \,= \,   
          \frac{ 1}{ 1 + {\dot \Sigma}^{BB}}
                        \Big ( 1+ \Sigma^{AB(1)} \Big)^2 \,, 
  \end{equation}   
  \begin{equation} \label{c31}
     V  : \quad  2\, F^{BBA} \,= \, \, F^{AAA}    
    \frac{ 1 + {\dot \Sigma}^{BB}}{ 1+ \Sigma_{\rm trans}} \,\, \,
        \longrightarrow  F^{hBA}_1\, \longrightarrow \, F^{AAh(1)}
   \quad {\rm due \,to}\quad (\ref{c11}),(\ref{c14}), 
   \end{equation}
  \begin{equation} \label{c32}
  VII_a  :  \qquad \Big (\frac{M}{m} \Big )^2 + \Sigma^{hh(2)} 
          \, = \, \frac{4}{g} \, F^{BBh(1)}\, \frac{R_4}{R_3}  \,,    
    \end{equation}
     \begin{equation} \label{c33}
   VII_{b+c}  :  \qquad  1 + {\dot \Sigma}^{hh} \, = \,
               ( 1 + {\dot \Sigma}^{BB}) \,\frac{ R_5}{R_3} \,.
  \end{equation} 

Resuming  the following task has been achieved: we first 
 treated the functional $\,\Ga^{\Lz} \,$ and its ancillary
  functionals $\, \Ga^{\Lz}_{\ga_{\tau}} , \Ga^{\Lz}_{\om}\,$
  with a BRS-insertion, disregarding the STI. There appear
  $ 37 + 7 $ relevant parameters.  Fixing  among these parameters
  a priori $ \, \kappa = 0\,$ (no tadpoles) 
  and $ \,\Sigma_{\rm long} = 0 \,$ (due to the field equation 
 of the antighost), and regarding the set (\ref{c19}) without  
 $ \Sigma_{\rm long} \,$, as  \emph{ renormalization
 constants to be chosen freely}, we can uniquely determine the
 remaining relevant parameters upon requiring the relevant part of
 the functional $\, \Ga^{\Lz}_1 \,$ to vanish, (\ref{ma6}), on 
account of the VSTI (\ref{vs22}). Finally, since the relevant
 part of the functional $\, \Ga^{\Lz}_1 \,$ 
vanishes , due to Proposition 3,
 (\ref{bdf}), its irrelevant part vanishes in the limit 
  $\, \Lambda_0 \to \infty \,$, too. Thus perturbatively
 the functional  $\, \Ga^{ 0, \, \infty} \,$ 
 and its ancillary funtionals 
 $\, \Ga^{0,\, \infty}_{\ga_{\tau}} , \Ga^{0,\, \infty}_{\om}\,$
 are finite and satisfy the STI, i.e. equation (\ref{vs22}) for
 $\, \Lambda_0 \to \infty \, $ with the r.h.s. vanishing.
\\[.5cm]
{\it Acknowledgement}~:\\[.1cm]
Both authors have been  lecturing at ESI, Vienna, about the subject
of this paper~; the ensuing discussions were important    
for its genesis; hospitality of ESI is therefore gratefully 
acknowledged.

\subsection*{Appendix A}
The bare functional $ L^{\Lambda_0, \Lambda_0} $ and the relevant
 part of the generating functional $\Gamma ^{ 0, \Lambda_0} $ for
 the proper vertex functions have the same general form. We present the
 latter and give the tree order explicitly. At the end we
 state the modification 
 to obtain the bare functional $ L^{\Lambda_0, \Lambda_0} $.
 Writing
 $$
 \Gamma^{\Lz} (\underline{A},\underline{h},\underline{B},\underline{\bar{c}},
\underline{c})   =\sum^4_{|n|=1} \Gamma_{|n|} + \Gamma_{(|n| > 4)}\, , 
 $$
 where $|n|$ counts the number of fields, we extracted the relevant part,
 i.e. its local field content with mass dimension not greater than four. 
 Moreover, in the sequel we do not underline the field variables though all 
 arguments in the $\Gamma $- functional should appear underlined, of course.

 \noindent
1) One-point function\\
$$ \Gamma_1 = \kappa \hat{h}(0). $$

\noindent
2) Two-point functions \\
\begin{eqnarray*}
\Gamma_2 =  \int_p \Big\{ \frac12 \,A^a_{\mu}(p) A^a_{\nu}(-p)
 \Gamma^{AA}_{\mu\nu}(p) +
 \frac12 \,h(p) h(-p) \Gamma^{hh}(p)  +
  \frac12\, B^a(p)B^a(-p) \Gamma^{BB}(p) \\
 - \bar{c}^a(p) c^a(-p) \Gamma^{\bar{c}c}(p)
 + A^a_{\mu}(p) B^a(-p) \Gamma^{AB}_{\mu}(p) \Big\}, \qquad \qquad \qquad 
 \end{eqnarray*}
 $$
 \Gamma^{AA}_{\mu\nu}(p) =  \delta_{\mu\nu}(m^2+\delta m^2) + 
 (p^2\delta_{\mu\nu}-p_{\mu}p_{\nu}) (1 + \Sigma_{\rm trans} (p^2)) 
  + \frac{1}{\alpha} p_{\mu}p_{\nu} (1 + \Sigma_{\rm long} (p^2)), $$  
 \begin{eqnarray*}
\Gamma^{hh}(p) &=& p^2 + M^2 + \Sigma^{hh}(p^2), \quad 
\Gamma^{BB}(p) = p^2 + \alpha m^2 + \Sigma^{BB}(p^2), \\
\Gamma^{\bar{c}c}(p) &=& p^2 + \alpha m^2 + \Sigma^{\bar{c}c}(p^2), \quad
\Gamma^{AB}_{\mu}(p) = ip_{\mu} \Sigma^{AB}(p^2).  
 \end{eqnarray*} 
Besides the unregularized tree order explicitly stated, 
there emerge 10 relevant parameters from the various self-energies:
$$\delta m^2 \,, \Sigma_{\rm trans}(0) \,, \Sigma_{\rm long}(0)\,,
    \Sigma^{hh}(0) \,, \dot{\Sigma}^{hh}(0)\, ,
  \Sigma^{BB}(0) \,, \dot{\Sigma}^{BB}(0) \,, \Sigma^{\bar{c}c}(0) \,,
 \dot{\Sigma}^{\bar{c}c}(0) \, ,\Si^{AB}(0) \, , $$   
where the notation
$\dot{\Si}(0) \equiv (\partial_{ p^2 }\Si)(0)$ has been used.
 We note that because of the regularization, the inverse of the
 regularized propagators (\ref{f7}) actually appears as the tree 
  order $(l=0)$ of the $2$-point functions. Due to the property (\ref{f6}),
  however, the regularizing factor $(\sigma _{0, \Lambda_0}(p^2) )^{-1}$   
 does not contribute to the relevant part.\\
 
\noindent
3) Three-point functions \\
We only present the relevant part explicitly. A relevant parameter 
vanishing in the tree order is denoted by $r \in \mathcal{O}(\hbar)$,
 otherwise it is denoted by $F$. Moreover, we
indicate an irrelevant part by a symbol ${\mathcal O}_n, \;
 n \in \mathbf{N}$, reminding that this
part vanishes like an $n$-th power of a momentum 
 when all momenta tend to zero homogeneously. 
\begin{eqnarray*}
\Gamma_3 &=& \int_p\int_q \big\{ \epsilon^{rst} A^r_{\mu}(p) A^s_{\nu}(q)
    A^t_{\lambda}(-p-q)
\Gamma^{AAA}_{\mu\nu\lambda}(p,q)   \\
&+ &  A^r_{\mu}(p) A^r_{\nu}(q) h(-p-q) \Gamma^{AAh}_{\mu\nu}(p,q) \\
&+ & \epsilon^{rst} B^r(p) B^s(q) A^t_{\mu}(-p-q) \Gamma^{BBA}_{\mu}(p,q) \\
& +&  h(p) B^r(q) A^r_{\mu}(-p-q) \Gamma^{hBA}_{\mu}(p,q) 
+ \epsilon^{rst} \bar{c}^r(p) c^s(q) A^t_{\mu}(-p-q) 
\Gamma^{\bar{c}cA}_{\mu}(p,q) \\
& +&  B^r(p) B^r(q) h(-p-q) \Gamma^{BBh}(p,q) + h(p) h(q) h(-p-q)
 \Gamma^{hhh}(p,q) \\
& +&  \bar{c}^r(p) c^r(q) h(-p-q) \Gamma^{\bar{c}ch}(p,q) 
+ \epsilon^{rst} \bar{c}^r(p) c^s(q) B^t(-p-q)
 \Gamma^{\bar{c}cB}(p,q) \big \},
\end{eqnarray*}

\begin{eqnarray*} \begin{array}{llllll}
\Gamma^{AAA}_{\mu\nu\lambda}(p,q) &=& \delta_{\mu\nu} i(p-q)_{\lambda} 
F^{AAA} + {\cal O}_3, \quad &  F^{AAA} &=& - \frac12 g + r^{AAA}, \\
\Gamma^{AAh}_{\mu\nu}(p,q) &=& \delta_{\mu\nu} F^{AAh} + {\cal O}_2, &
F^{AAh} &=&  \frac12 mg + r^{AAh}, \\
\Gamma^{BBA}_{\mu}(p,q) &=& i(p-q)_{\mu} F^{BBA} + {\cal O}_3, &
F^{BBA} &=&  - \frac14 g + r^{BBA}, \\
\Gamma^{hBA}_{\mu}(p,q) &=& i(p-q)_{\mu} F_1^{hBA} &
F_1^{hBA} &=&  \frac12 g + r_1^{hBA}, \\
& & + i(p+q)_{\mu} r^{hBA}_2 + {\mathcal O}_3,\\
\Gamma^{\bar{c}cA}_{\mu}(p,q) &=& ip_{\mu} F_1^{\bar{c}cA} +
 iq_\mu r^{\bar{c}cA}_2 
+ {\cal O}_3, &
F_1^{\bar{c}cA} &=& g + r_1^{\bar{c}cA}, \\
\Gamma^{BBh}(p,q) &=& F^{BBh} + {\mathcal O}_2, &
F^{BBh} &=&  \frac14 g \frac{M^2}{m} + r^{BBh}, \\
\Gamma^{hhh}(p,q) &=& F^{hhh} + {\mathcal O}_2, &
F^{hhh} &=&  \frac14 g \frac{M^2}{m} + r^{hhh}, \\
\Gamma^{\bar{c}ch}(p,q) &=& F^{\bar{c}ch} + {\mathcal O}_2, &
F^{\bar{c}ch} &=&  - \frac12 \alpha gm + r^{\bar{c}ch}, \\
\Gamma^{\bar{c}cB}(p,q) &=& F^{\bar{c}cB} + {\mathcal O}_2, &
F^{\bar{c}cB} &=&  \frac12 \alpha gm + r^{\bar{c}cB}. 
\end{array}
\end{eqnarray*}
The 3-point functions $AAB$ and $BBB$ have no relevant local content. \\
\noindent
4) Four-point functions \\
Defining as before parameters $r$ and $F$, then 
\begin{eqnarray*}
\Gamma_4|_{\rm rel} &=& \int_k \int_p \int_q \big\{ \epsilon^{abc}
 \epsilon^{ars} A^b_{\mu}(k)
A^c_\nu(p)A^r_\mu(q) A^s_\nu(-k-p-q) F_1^{AAAA} \\
& & + A^r_\mu(k) A^r_\mu(p) A^s_\nu(q) A^s_\nu(-k-p-q) r_2^{AAAA} \\
& & + A^a_\mu(k)A^b_\mu(p) \bar{c}^r(q) c^s(-k-p-q) (\delta^{ab}
   \delta^{rs}r_1^{AA\bar{c}c}
+ \delta^{ar}\delta^{bs}r_2^{AA\bar{c}c}) \\
& & + A^a_\mu(k)A^b_\mu(p) B^r(q) B^s(-k-p-q)
 (\delta^{ab}\delta^{rs}F_1^{AABB}
+ \delta^{ar}\delta^{bs}r_2^{AABB}) \\
& & + B^a(k)B^b(p) \bar{c}^r(q) c^s(-k-p-q)
 (\delta^{ab}\delta^{rs}r_1^{BB\bar{c}c}
+ \delta^{ar}\delta^{bs}r_2^{BB\bar{c}c}) \\
& & + h(k)h(p)h(q)h(-k-p-q) F^{hhhh} \\
& & + B^r(k)B^r(p)h(q)h(-k-p-q) F^{BBhh} \\
& & + B^r(k)B^r(p)B^s(q)B^s(-k-p-q) F^{BBBB} \\
& & + A^r_\mu(k)A^r_\mu(p)h(q)h(-k-p-q) F^{AAhh} \\
& & + h(k)h(p)\bar{c}^r(q)c^r(-k-p-q) r^{hh\bar{c}c} \\
& & + \bar{c}^a(k)c^a(p)\bar{c}^r(q)c^r(-k-p-q) r^{\bar{c}c\bar{c}c} \\
& & + \epsilon^{rst}h(k)B^r(p)\bar{c}^s(q)c^t(-k-p-q) r^{hB\bar{c}c} \big\},
\end{eqnarray*} 
\begin{eqnarray*}\begin{array}{llllll}
F_1^{AAAA} &=& \frac14 g^2 + r_1^{AAAA}, \quad &
F_1^{AABB} &=& \frac18 g^2 + r_1^{AABB}, \\
F^{hhhh} &=& \frac{1}{32} g^2 \left( \frac Mm \right)^2 + r^{hhhh}, &
F^{BBhh} &=& \frac{1}{16} g^2 \left( \frac Mm \right)^2 + r^{BBhh}, \\
F^{BBBB} &=& \frac{1}{32} g^2 \left( \frac Mm \right)^2 + r^{BBBB}, &
F^{AAhh} &=& \frac18 g^2 + r^{AAhh}.
\end{array} 
\end{eqnarray*}
Hence, $\Gamma^{\Lz} $ in total involves $1 + 10 + 11 + 15 = 37$ relevant
 parameters. \\
 We now obtain the form of the bare functional $ L^{\Lambda_0, \Lambda_0}$, 
 together with its order $l=0$ explicitly given, 
 upon deleting in the two-point functions the contributions of the 
 order $l=0$, i.e. keeping there only the 10 parameters which 
appear in the various self-energies.
\subsection*{Appendix B}
 Analysing the STI, vertex functions (\ref{vs21})
 with one operator insertion, 
 generated by the BRS-variations, have to be considered, too.
 These insertions have mass dimension $D=2$. We remind the notation
  (\ref{i7}) and (\ref{i8}) of the corresponding Fourier-transform, 
  presenting the respective relevant part of these four vertex functions 
 with one insertion,
\begin{eqnarray*}
\hat{\Gamma}^{\Lz}_{\gamma^a_\mu}(q, \uFi)|_{\rm rel} &=& 
               - \,iq_{\mu}\, \uca (-q)\,R_1 
   + \epsilon^{arb} \int_k {\uA}^r_\mu(k) {\uc}^b(-q-k) g R_2 \, ,\\
 \hat{\Gamma}^{\Lz}_{\gamma}(q; \uFi)|_{\rm rel} &=& 
        - \,\frac12 \,g  \int_k {\uB}^r(k) \,{\uc}^r(-q-k) R_3 \,, \\
 \hat{\Gamma}^{\Lz}_{\gamma^a}(q; \uFi)|_{\rm rel} &=& 
     m \uca(-q) R_4 \\ 
  & &+ \int_k {\uh}(k)c^a(-q-k)\frac12\,g R_5 
  + \epsilon^{arb} \int_k {\uB}^r(k)\,{\uc}^b (-q-k) \frac12\,g R_6 \, ,\\
 \hat{\Gamma}^{\Lz}_{\omega^{\,a}}(q; \uFi)|_{\rm rel} &=& 
      \epsilon^{ars} \int_k {\uc}^r(k)\,{\uc}^s(-q-k) \frac12 \,g R_7.
\end{eqnarray*} 
There appear 7 relevant parameters 
$$R_i = 1 + r_i \, , \qquad r_i = {\cal O}(\hbar), \qquad i = 1,...,7. $$
All the other two-point functions, and the higher ones, of course,
 are of irrelevant type.
 \subsection*{Appendix C}
As a consequence of the expansion in the mass parameters
the conditions following from the fact that the 
relevant part of the functional $\Ga_1\,$ should vanish 
\[ 
\Gamma_1(\uA,\uh,\uB,\ubc,\uc)|_{\dim \le 5} \begin{array}{c}
 ! \\ = \\ \\
\end{array} 0 .
\]
can be reordered according to the value of
$\nu\,$ which appears. We get contributions for $0 \le \nu \le 3\,$. 
The value of $\, \nu\,$ in the various relevant couplings is indicated
 as a superscript in parentheses {\it if $\nu >0\,$}.  
We explicitly indicate the momentum and the power of $m\,$ in front of
each STI. The power of $m\,$ indicates the value of $\nu\,$
in the corresponding contribution to $\Ga_1\,$.\\[0.2cm]
\noindent
Two fields
\begin{enumerate}
\item[I)] $\; \delta_{A^a_\mu(q)} \delta_{c^r(k)} \Gamma_1|_0$
\begin{enumerate}
\item[a) $0 \, \stackrel {!} {= } \,$]$   m^2\, q_\mu \left\{ -
    (1+\de m^2_{\,(2)} )R_1 + \sum^{AB(1)}  R_4 + 
1 + \frac{1}{\alpha} \sum^{\bar{c}c(2)} \right\} $,
\item[b) $0  \, \stackrel {!} {= } \,$]$  q^2 q_\mu \Big\{ - {1\over
      \al}(1 + \sum_{\rm long} ) R_1 
+{1\over  \al} (1 + \dot{\sum}^{\bar{c}c}) 
\Big\} $.
\end{enumerate}
\end{enumerate}  
\begin{enumerate}
\item[II)] $\; \delta_{B^a(q)} \delta_{c^r(k)} \Gamma_1|_0$
\begin{enumerate}
\item[a) $0  \, \stackrel {!} {= }$]$ m^3 \left\{(\alpha +\sum^{BB(2)})R_4
- (\alpha + \sum^{\bar{c}c(2)}) 
          - \frac{g}{2}\, \ka^{(3)} R_3 \, \right\}$.
\item[b) $0   \stackrel {!} {=}$]$m\,q^2  \Big\{ - \sum^{AB(1)} R_1
+ (1 +\dot{\sum}^{BB})R_4- (1 + \dot{\sum}^{\bar{c}c}) \Big\} $.
\end{enumerate}
\end{enumerate}
\vspace{0.2cm}
\noindent
Three fields
\begin{enumerate}
\item[III)] $\; \delta_{A^r_\mu(p)} \delta_{A^s_\nu(q)} \delta_{c^t(k)}
\Gamma_1|_0$
\begin{enumerate}
\item[a) $ 0    \stackrel {!} {= }($]$\!\!\!p_\mu p_\nu - q_\mu q_\nu)\!
\Big\{\!\! -\! 2 F^{AAA}R_1\! - \! \frac{1}{\alpha}
(F_1^{\bar{c}cA}\! - r_2^{\bar{c}cA})\! +\!
\left[\frac{1}{\alpha} ( 1\! +
\sum_{\rm long})\!-\!
(1\!+\sum_{\rm trans})\right]gR_2\Big\},$
\item[b) $0   \stackrel {!} {= }    ($]$\!\! p^2-q^2)\delta_{\mu\nu}
\left\{ 2 F^{AAA} R_1 +  (1 + \sum_{\rm trans} ) g R_2
\right\} $,
\end{enumerate}
\item[IV)] $\; \delta_{A^r_\mu(p)} \delta_{B^s(q)} \delta_{c^t(k)}
\Gamma_1|_0$
\begin{enumerate}
\item[a) $0   \, \stackrel {!} {= } \,$]$   m\, p_\mu \left\{ 2F^{BBA} R_4 +
\frac12 g \sum^{AB(1)} R_6 + \frac{1}{\alpha}
F^{\bar{c}cB, (1)} -  r_2^{\bar{c}cA}\right\}$,
\item[b) $0   \, \stackrel {!} {= } \,$]$ m\, q_\mu 
\left\{ g \sum^{AB(1)} R_2
+ 4F^{BBA} R_4 +  (F_1^{\bar{c}cA} - r_2^{\bar{c}cA}) \right\}$,
\end{enumerate}
\item[V)] $ \delta_{B^r(p)} \delta_{B^s(q)} \delta_{c^t(k)} \Gamma_1|_0
$
\begin{enumerate}
\item[] $\!\!\!\!\!\!\!\!\!\!\!\!\!\! \!\!
0 \! \stackrel {!} {= }\! (p^2\!-q^2)\!\left\{ 2R_1 F^{BBA} +
(1 + \dot{\sum}^{BB}) {g \over 2} R_6 
\right\} $,
\end{enumerate}
\item[VI)] $\; \delta_{A^r_\mu(p)} \delta_{h(q)} \delta_{c^t(k)}
\Gamma_1|_0$
\begin{enumerate}
\item[a) $0 \, \stackrel {!} {= }\, $]$   m\, p_\mu \Big\{ -2R_1F^{AAh(1)}+
R_4 (F_1^{hBA} - r_2^{hBA}) 
+ \sum^{AB(1)} \frac12 gR_5 - \frac{1}{\alpha} F^{\bar{c}ch(1)} \Big\}$,
\item[b) $0   \, \stackrel {!} {= }
 \,$]$ m\, q_\mu \left\{ -2 R_1F^{AAh(1)} +
2R_4  F_1^{hBA}\right\}$,
\end{enumerate}
\item[VII)] $\; \delta_{h(p)} \delta_{B^s(q)} \delta_{c^t(k)}
\Gamma_1|_0$
\begin{enumerate}
\item[a) $0   \, \stackrel {!} {= } \,$]$   m^2\left\{
(\frac{M^2}{m^2} + \sum^{hh(2)}) (-
\frac12 g R_3) + 2 F^{BBh(1)} R_4 + F^{\bar{c}ch(1)}
 + (\alpha+ \sum^{BB(2)}) \frac12 g R_5\right\} $,
\item[b) $0   \, \stackrel {!} {= } \,$]$ p^2 \Big\{ F_1^{hBA}R_1 - (1 +
   \dot{\sum}^{hh}) \frac12 gR_3 
\Big\}$,
\item[c) $0   \, \stackrel {!} {= } \,$]$ q^2 \Big\{- F^{hBA}_1R_1 + (1 +
  \dot{\sum}^{BB}) \frac12 gR_5 
\Big\}$,
\item[d) $0   \, \stackrel {!} {= } \,$]$ k^2 \left\{ r_2^{hBA} R_1
\right\}$,
\end{enumerate}   
\item[VIII)] $\; \delta_{c^t(q)} \delta_{c^s(p)} \delta_{\bar{c}^r(k)}
\Gamma_1|_0$
\begin{enumerate}
\item[a) $0   \, \stackrel {!} {= } \,$]$   m^2 \left\{2
    F^{\bar{c}cB(1)}R_4 - (\alpha  +  \sum^{\bar{c}c (2)})  g R_7\right\}$,
\item[b) $0   \, \stackrel {!} {= } \,$]$ k^2 \Big\{ F_1^{\bar{c}cA}R_1 -
r_2^{\bar{c}cA}R_1 - (1 + \dot{\sum}^{\bar{c}c}) gR_7 \Big\}$,
\item[c) $0   \, \stackrel {!} {= } \,$]$ (p^2 + q^2) \Big\{
r_2^{\bar{c}cA}R_1 \Big\}$.
\end{enumerate}
\end{enumerate}
\noindent 
Four fields
\begin{enumerate}
\item[IX)] $\; \delta_{h(p)} \delta_{h(q)} \delta_{B^1(k)}
\delta_{c^1(l)} \Gamma_1|_0$ \\
\\
$0   \, \stackrel {!} {= } \, m\,\left\{
6 F^{hhh, (1)} (- \frac12 gR_3) + 4 F^{BBhh} 
R_4 + 2 F^{BBh, (1)} g R_5 + 2 r^{hh\bar{c}c}\right\} $.
\item[X)] $\; \delta_{B^1(k)} \delta_{B^1(p)} \delta_{B^2(q)}
\delta_{c^2(l)} \Gamma_1|_0$ \\
\\
$0   \, \stackrel {!} {= } \, m\left\{- F^{BBh, (1)} gR_3 + 8F^{BBBB}  R_4
 +  \left( 2r_1^{BB\bar{c}c} +r^{BB\bar{c}c}_2 \right)\right\} $.
\item[XI)] $\; \delta_{h(l)} \delta_{\bar{c}^3(k)} \delta_{c^1(p)}
\delta_{c^2(q)} \Gamma_1|_0$ \\
\\
$0   \, \stackrel {!} {= } \,  m\left\{2r^{hB\bar{c}c}  R_4 
+ F^{\bar{c}cB (1)} gR_5 + F^{\bar{c}ch, (1)} gR_7 \right\} $.
\item[XII)] $\; \delta_{c^2(k)} \delta_{\bar{c}^2(l)} \delta_{c^1(p)}
\delta_{B^1(q)} \Gamma_1|_0$ \\
\\
$0   \, \stackrel {!} {= } \,  m\left\{ F^{\bar{c}ch(1)} (- \frac12 g R_3) +
(2r_1^{BB\bar{c}c}- r_2^{BB\bar{c}c}) R_4 
 + F^{\bar{c}cB(1)} (\frac12 gR_6 - gR_7) + 2
 r^{\bar{c}c\bar{c}c}\right\}   $.
\item[XIII)$_1$] $\; \delta_{A^1_\mu(k)} \delta_{A^2_\nu(p)}
\delta_{B^1(q)} \delta_{c^2(l)} \Gamma_1|_0$ \\
\\
$0   \, \stackrel {!} {= } \,  2r_2^{AABB} R_4 + r_2^{AA\bar{c}c} $.
\item[XIII)$_2$] $\; \delta_{A^1_\mu(k)} \delta_{A^1_\nu(p)}
\delta_{B^2(q)} \delta_{c^2(l)} \Gamma_1|_0$ \\
\\
$0   \, \stackrel {!} {= } \, 
\,  m\left\{ -F^{AAh(1)}  g R_3 +
4F_1^{AABB} R_4 
 + 2  r_1^{AA\bar{c}c}\right\}   $.
\item[XIV)] $\; \delta_{A^1_\mu(p)} \delta_{A^1_\nu(q)}
\delta_{A^2_\rho(k)} \delta_{c^2(l)} \Gamma_1|_0$
\begin{enumerate}
\item[a)] $0   \, \stackrel {!} {= } \, 2 \delta_{\mu\nu} l_\rho \Big\{
4 (F^{AAAA}_1 + r_2^{AAAA}) R_1 
+ 2F^{AAA} gR_2 + \frac{1}{\alpha} r_1^{AA\bar{c}c}
\Big\}$,
\item[b)] $0   \, \stackrel {!} {= } \, \delta_{\mu\nu} (p_\rho +
q_\rho)  \left\{ \frac{2}{\alpha} r_1^{AA\bar{c}c}
\right\}$,
\item[c)] $0   \, \stackrel {!} {= } \, (\delta_{\mu\rho}l_\nu +
\delta_{\nu\rho}l_\mu) \left\{ - 4 F_1^{AAAA}R_1 - 2 F^{AAA} gR_2 
\right\}$,
\item[d)] $0   \, \stackrel {!} {= } \, (\delta_{\mu\rho}p_\nu +
\delta_{\nu\rho}q_\mu) \left\{ 0 \right\}$,
\item[e)] $0   \, \stackrel {!} {= } \, (\delta_{\mu\rho}q_\nu +
\delta_{\nu\rho}p_\mu)  \left\{ - \frac{1}{\alpha} r_2^{AA\bar{c}c}\right\}$.
\end{enumerate}   
\item[XV)$_1$] $\; \delta_{B^1(p)} \delta_{B^1(q)} \delta_{A^2_\mu(k)}
\delta_{c^2(l)} \Gamma_1|_0$
\begin{enumerate}
\item[a)] $0   \, \stackrel {!} {= } \, l_\mu  \Big\{ 4 F_1^{AABB}R_1 +
2F^{BBA} gR_6 \Big\}$,
\item[b)] $0   \, \stackrel {!} {= } \, k_\mu 
\left\{ r_1^{BB\bar{c}c}\right\}$,
\end{enumerate}
\item[XV)$_2$] $\; \delta_{B^1(p)} \delta_{B^2(q)} \delta_{A^1_\mu(k)}
\delta_{c^2(l)} \Gamma_1|_0$
\begin{enumerate}
\item[a)] $0   \, \stackrel {!} {= } \, p_\mu  \left\{ -2r_2^{AABB}R_1 +
2F^{BBA} gR_2 + F_1^{hBA} gR_3 \right\}$,
\item[b)] $0   \, \stackrel {!} {= } \, q_\mu \left\{-2 r_2^{AABB}R_1 -
2F^{BBA} gR_2 + 2F^{BBA}gR_6 \right\}$,
\item[c)] $0   \, \stackrel {!} {= } \, k_\mu \Big\{-2 r_2^{AABB}R_1 +
F_1^{hBA} \frac12 gR_3 + r_2^{hBA}\frac12 gR_3
 + F^{BBA} gR_6 - \frac{1}{\alpha} r_2^{BB\bar{c}c}
\Big\}$,
\end{enumerate}
\item[XVI)] $\; \delta_{h(p)} \delta_{A_{\mu}^1(k)} \delta_{B^2(q)}
\delta_{c^3(l)} \Gamma_1|_0$
\begin{enumerate}
\item[a)] $0   \, \stackrel {!} {= } \, p_\mu  \left\{ F_1^{hBA}g(R_6
-R_2) - r_2^{hBA} gR_2 \right\}$,
\item[b)] $0   \, \stackrel {!} {= } \, q_\mu \left\{F_1^{hBA}gR_2 -
r_2^{hBA} gR_2 + 2F^{BBA}gR_5 \right\}$,
\item[c)] $0   \, \stackrel {!} {= } \, k_\mu \Big\{F_1^{hBA} \frac12
gR_6 - r_2^{hBA} \frac12 gR_6 + F^{BBA} gR_5 
 - \frac{1}{\alpha} r^{hB\bar{c}c} \Big\}$,
\end{enumerate}  
\item[XVII)] $\; \delta_{h(p)} \delta_{h(q)} \delta_{A^1_\mu(k)}
\delta_{c^1(l)} \Gamma_1|_0$
\begin{enumerate}
\item[a)] $0   \, \stackrel {!} {= } \, l_\mu  \left\{ 4F^{AAhh}R_1 -
F_1^{hBA} gR_5 \right\}$,
\item[b)] $0   \, \stackrel {!} {= } \, k_\mu \left\{r_2^{hBA}gR_5 +
\frac{2}{\alpha} r^{hh\bar{c}c}  \right\}$.
\end{enumerate}
\item[XVIII)] $\; \delta_{A^2_\mu(k)} \delta_{c^2(p)} \delta_{c^1(q)}
\delta_{\bar{c}^1(l)} \Gamma_1|_0$
\begin{enumerate}
\item[a)] $0   \, \stackrel {!} {= } \, l_\mu  \left\{
F_1^{\bar{c}cA}g(R_2 -R_7) + \frac{2}{\alpha} r^{\bar{c}c\bar{c}c}\right\}$,
\item[b)] $0   \, \stackrel {!} {= } \, p_\mu \Big\{
2r_1^{AA\bar{c}c}R_1 + r_2^{\bar{c}cA} g(R_2 - R_7) $
$ + \frac{2}{\alpha} r^{\bar{c}c\bar{c}c} \Big\} $,
\item[c)] $0   \, \stackrel {!} {= } \, q_\mu \Big\{ -r_2^{AA\bar{c}c}
R_1 - r_2^{\bar{c}cA} gR_7  + \frac{2}{\alpha} r^{\bar{c}c\bar{c}c}\Big\}$.
\end{enumerate}
\end{enumerate}
\vspace{0.2cm}
\noindent
Five fields
\begin{enumerate}
\item[XIX)] $\; \delta_{h(p)} \delta_{h(q)} \delta_{h(k)}
\delta_{B^1(l)} \delta_{c^1(l^\prime)} \Gamma_1|_0$ \\
\\
$0   \, \stackrel {!} {= } \, -2F^{hhhh}R_3 + F^{hhBB} R_5
$.
\item[XX)] $\; \delta_{h(p)} \delta_{B^1(q)} \delta_{B^1(k)}
\delta_{B^2(l)} \delta_{c^2(l^\prime)} \Gamma_1|_0$ \\
\\
$0   \, \stackrel {!} {= } \, -F^{BBhh} R_3 + 2F^{BBBB} R_5
$.
\item[XXI)] $\; \delta_{A^1_\mu(k)} \delta_{A^1_\nu(p)} \delta_{h(k)}
\delta_{B^2(l)} \delta_{c^2(l^\prime)} \Gamma_1|_0$ \\
\\
$0   \, \stackrel {!} {= } \, -F^{AAhh} R_3 + F_1^{AABB} R_5
$.
\item[XXII)] $\; \delta_{A^1_\mu(k)} \delta_{B^1(p)}
\delta_{c^1(l^\prime)} \delta_{A^2_\nu(q)} \delta_{B^3(l)} \Gamma_1|_0$
\\
\\
$0   \, \stackrel {!} {= } \, r_2^{AABB} (R_6- 2R_2) $.
\item[XXIII)] $\; \delta_{A^1_\mu(k)} \delta_{B^1(q)}
\delta_{A^2_\nu(p)} \delta_{c^2(l^\prime)} \delta_{h(l)} \Gamma_1|_0$ 
\\
\\
$0   \, \stackrel {!} {= } \, r_2^{AABB} R_5  $.
\item[XXIV)] $\; \delta_{A^3_\mu(k)} \delta_{A^3_\nu(p)}
\delta_{\bar{c}^2(q)} \delta_{c^3(l)}  \delta_{c^1(l^\prime)}
\Gamma_1|_0$ \\
\\
$0   \, \stackrel {!} {= } \, r_2^{AA\bar{c}c} R_2  + r_1^{AA\bar{c}c}
R_7 $.   
\item[XXV)] $\; \delta_{A^3_\mu(k)} \delta_{\bar{c}^3(q)}
\delta_{A^2_\nu(p)}  \delta_{c^3(l)} \delta_{c^1(l^\prime)}
\Gamma_1|_0$ \\
\\
$0   \, \stackrel {!} {= } \, r_2^{AA\bar{c}c} (3R_2 - R_7)
$.
\item[XXVI)] $\; \delta_{B^1(p)} \delta_{B^1(q)} \delta_{\bar{c}^1(k)}
\delta_{c^2(l)}  \delta_{c^3(l^\prime)}  \Gamma_1|_0$ \\
\\
$0   \, \stackrel {!} {= } \, r_2^{BB\bar{c}c} (R_6 - R_7)  -
r_1^{BB\bar{c}c} R_7 $.
\item[XXVII)] $\; \delta_{B^1(p)}  \delta_{\bar{c}^1(k)} \delta_{B^2(q)}
\delta_{c^3(l)}  \delta_{c^1(l^\prime)}  \Gamma_1|_0$ \\
\\
$0   \, \stackrel {!} {= } \, -r^{hB\bar{c}c} R_3  + r_2^{BB\bar{c}c}
(3R_6 - 2R_7) $.
\item[XXVIII)] $\; \delta_{h(p)}  \delta_{h(q)} \delta_{\bar{c}^1(k)}
\delta_{c^2(l)} \delta_{c^3(l^\prime)}   \Gamma_1|_0$ \\
\\
$0   \, \stackrel {!} {= } \, r^{hB\bar{c}c} R_5  + r^{hh\bar{c}c} R_7$.
\item[XXIX)] $\; \delta_{h(p)}  \delta_{B^1(q)} \delta_{c^1(l)}
\delta_{\bar{c}^2(k)} \delta_{c^2(l^\prime)}   \Gamma_1|_0$ \\
\\
$0   \, \stackrel {!} {= } \, 2r^{hh\bar{c}c} R_3  - 2r_1^{BB\bar{c}c} R_5
+ r_2^{BB\bar{c}c} R_5 + r^{hB\bar{c}c} (-R_6 + 2R_7) $.
\end{enumerate}
\newpage            
     
\noindent
{\bf References}
 
\begin{itemize}
\item[{[BRS]}]  C. Becchi, A. Rouet and R. Stora, Renormalization of
        gauge theories, \\ Ann. Phys. (N.Y.) {\bf 98}, 287 - 321 (1976). 
\item[{[FS]}] L.D. Faddeev, A.A. Slavnov, \emph{Gauge Fields: Introduction
             to Quantum Theory},\\ Benjamin, Reading MA 1980. 
\item[{[KK]}] G. Keller, Ch. Kopper: Perturbative Renormalization of Composite
Operators via Flow Equations I. Commun. Math. Phys. {\bf 148}, 445-467 (1992)  

\item[{[KKS]}] G. Keller, Ch. Kopper, M. Salmhofer: 
Perturbative Renormalization
and Effective Lagrangians in $\Phi^4_4$. Helv. Phys. Ac\-ta {\bf 65}, 32-52
(1991).
\item[{[KM]}] Ch. Kopper, V.F. M\"uller, Renormalization Proof for
      Spontaneously Broken Yang-Mills Theory with Flow Equations,
      Commun. Math. Phys. {\bf 209}, 477 - 516 (2000).  
\item[{[KMR]}] Ch. Kopper, V.F. M\"uller and Th. Reisz,
Temperature Independent Renormalization of Finite Temperature 
Field Theory, Ann. Henri Poincar{\'e} {\bf 2}, 
387-402 (2001).
\item[{[M]}] V.F. M\"uller, Perturbative Renormalization by Flow 
    Equations,\\  Rev. Math. Phys. {\bf 15}, 491 - 558 (2005). 
\item[{[P]}] J. Polchinski: Renormalization and Effective Lagrangians,\\
Nucl.Phys. {\bf B231}, 269-295 (1984).
\item[{[T]}] I.V. Tyutin: Gauge Invariance In Field Theory And Statistical
      Mechanics, \\
  Lebedev FIAN 39 (1975).   
\item[{[W]}]  K. Wilson: Renormalization group and critical phenomena 
I. Renormalization group and the Kadanoff scaling picture,
Phys.Rev. {\bf B4}, 3174-3183 (1971),\\ 
K. Wilson: Renormalization group and critical pheno\-mena
II. Phase cell analysis of
critical behaviour, Phys.Rev. {\bf B4}, 3184-3205  (1971).
\item[{[WH]}] F. Wegner, A. Houghton: Renormalization Group Equations
  for Critical Phenomena, Phys. Rev. {\bf  A8}, 401-412 (1973). 
\item[{[Z]}] J. Zinn-Justin:  Quantum Field Theory and Critical
Phenomena, ch. 21, Clarendon Press,Ox\-ford, 3rd ed. 1997, and 
J. Zinn-Justin in: Trends in Elementary Particle Theory, Lecture Notes
in Physics {\bf 37}, 2 - 40, Springer-Verlag 1975. 
 \end{itemize}

 \end{document}